\documentclass[preprint]{aastex}
\slugcomment{submitted to {\it The Astronomical Journal}}
\begin{document}
\title{The $uvby$H$\beta$ Metallicity Calibration for G and K Dwarfs}
\author{Bruce A. Twarog, Luis C. Vargas, and Barbara J. Anthony-Twarog}
\affil{Department of Physics and Astronomy, University of Kansas, Lawrence, KS 66045-7582}
\affil{Electronic mail: btwarog@ku.edu, lcvargas@ku.edu, bjat@ku.edu}

\begin{abstract}
The metallicity dependence of the primary indices of the $uvby$ photometric system for cooler dwarfs ($T_e$ $\sim$ 6500 K to 5000K) is investigated. The data base for the analysis is composed of the overlap between a composite catalog of selected, high-dispersion spectroscopic abundances for 1801 stars on the metallicity scale of \citet{vf05} and a merged catalog of high-precision $uvby$H$\beta$ photometry for over 35,000 stars. While [Fe/H] for F dwarfs is best estimated from $m_1$, with a modest dependence on $c_1$ as expected, for hotter G dwarfs the pattern reverses and $c_1$ becomes the dominant index. For cooler G dwarfs and K stars, the $c_1$ dominance continues, but a discontinuity appears such that stars between $b-y$ = 0.50 and 0.58 with [Fe/H] $\geq$ +0.25 have $m_1$ and $c_1$ indices that classify them as subgiants, confirming an earlier result based upon a much smaller sample. The reversal in the sensitivity to $m_1$ and $c_1$ is traced, in part, to the metallicity sensitivity of the $b-y$ index.  Moreover, $b-y$ grows larger in a non-linear fashion for stars above solar metallicity, leading to an overestimate of the reddening for super-metal-rich stars from some standard intrinsic color relations. Based upon successful tests using indices from synthetic spectra and the empirical trends among the observations, metallicity calibrations tied to H$\beta$ rather than $b-y$ have been derived for [Fe/H] $\geq$ $-$1.0, generating dispersions among the residuals ranging from 0.061 dex to 0.085 dex over the entire temperature range of interest. The new calibrations have the added advantage of being significantly less sensitive to errors in reddening than previous calibrations.
\end{abstract}

\section{INTRODUCTION}
Since its introduction by \citet[and references therein]{st66} as a potential tool for studying stellar populations, the $uvby$ photometric system, supplemented by the reddening-free H$\beta$ index, has more than fulfilled its promise by becoming the premiere photometric approach to defining fundamental stellar parameters from effective temperature to metallicity to surface gravity over the majority of the color-magnitude diagram (CMD). Development and calibration of the system have progressed through publications too numerous to mention; a few of the more obvious references include \citet{cr75,cr78,cr79} for F, B, and A stars, respectively. Revisions and extensions to cooler dwarfs and more metal-deficient stars include \citet{o84,o88,n88,sn89}. Applications to cool giants of the disk and halo can be found in \citet{bo70,bo80,ri89,tat91,gr92,att94,hi00}. The sensitivity of the system has been enhanced by the addition of a fifth filter, $Ca$, centered on the H and K lines of calcium. By replacing the $v$ filter in the $m_1$ = $(v-b) - (b-y)$ index with the $Ca$ filter, we created a new, metallicity-sensitive $hk$ index \citep{att91,tat95,att98,at00}. It was, in fact, the need to revise a preliminary metallicity calibration of the $hk$ index \citep{at02, tw03} and the desire to tie it to the same metallicity scale as defined by the $m_1$ index that unexpectedly led to the current investigation.

The $hk$ system includes data for a few thousand stars, most of which have published $uvby$ data, with only a modest fraction included among high dispersion spectroscopic catalogs. To expand the potential sample of stars with abundances for defining the $hk$ calibration, the decision was made to use photometric metallicities tied to the $uvby$ system for stars with high-precision, published photometry. This exercise raised a variety of issues, not the least of which was an apparent flaw in the most commonly used metallicity calibration for G dwarfs \citep{sn89} that systematically underestimated the abundances of stars more metal-rich than the sun, as detailed in \citet{tw02}. Though the sample available for testing was small, for the cooler dwarfs it was noted that the photometric metallicity dependence on $c_1$ was seriously underestimated. In fact, the $c_1$ index grew so large for the most metal-rich stars of the sample that their indices resembled those for subgiants and giants  rather than unevolved dwarfs. Because of the value of identifying cooler, metal-rich dwarfs for followup observations in planetary searches, not to mention the statistical analyses of stellar populations in the solar neighborhood, these shortcomings could bias survey programs tied to $uvby$ photometry against the discovery of such objects. 

It should be remembered that a partial source of the ongoing problem was the application of the \citet{sn89} metallicity calibration, tied predominantly to stars of solar abundance or less, to more metal-rich stars where the number of calibrators was minimal. \citet{sn89} divided the sample into F stars and G stars, using 103 and 116 calibrators, respectively, for each group, with the spectroscopic abundances coming from a mixed sample of high dispersion spectroscopic studies with estimated typical errors in [Fe/H] above 0.10 dex for a single abundance determination. 

With the problem identified, a predictable solution was a rederivation of the metallicity calibration using an expanded sample of stars, as illustrated by the work of \citet{ma02}(MA).  Following a pattern set by \citet{sn89}, the revised calibration focused on the development of a universal function that permitted calculation of metallicity over the color range from $b-y$ = 0.29 to 0.57, [Fe/H] = $-$2.0 to +0.5, and $M_V$ $>$ 1.0, though the emphasis of the investigation was on the metal-rich end of the scale. The final sample of 633 stars made use of two composite catalogs, \citet{hm98} for photometry and \citet{ca01} for spectroscopy. While \citet{sn89} used two functions of 8 terms each to cover the hotter and cooler color ranges, respectively, MA used one function of 20 terms to calibrate their entire color range.

More recently, \citet{no04} (N04) have revisited the \citet{sn89} approach and revised the two calibrations for F and G stars, making use of 342 stars for the hotter group ($b-y$ between 0.18 and 0.38) and 72 stars for the cooler range ($b-y$ between 0.44 and 0.59), with a metallicity coverage from [Fe/H] = $-1.8$ to +0.8. The spectroscopic sample is based upon a mixture of mostly recent literature sources, while the photometry comes from their own homogeneous, high-precision catalog.

With the appearance of the spectroscopic catalog of \citet{vf05} (VF) , the evolution of photometric calibrations of stellar parameters enters a potentially unique phase. The VF sample includes derivation of the temperature, surface gravity, metallicity, and rotational velocities from echelle spectroscopy for 1039 F, G, and K dwarfs, plus the Sun, all reduced with a uniform procedure using fits to synthetic spectra. The homogeneity and size of the catalog remove one of the major weaknesses of past $uvby$ analyses, the need to merge mixed spectroscopic samples derived using variable techniques to obtain a set of calibrators large enough to map the entire parameter space of interest. In fact, as noted above for \citet{sn89}, many of the problems with earlier calibrations are tied to extrapolations of calibration functions to parameter ranges where they were inadequately tested or the spectroscopic abundances were poorly determined. With a large data base to work with, it now becomes easier to transform smaller spectroscopic surveys to a common system, testing for more corrections than a simple offset in the zero-points of the individual metallicity scales. 

Finally, with a large enough sample of stars, it is now possible to map out the varying sensitivity of the key indices to changes in temperature, surface gravity, and, most importantly for this study, metallicity. An example of the value of this exceptional sample can be found in \citet{att07}, where the breakdown of the traditional $b-y$, H$\beta$ relations of \citet{o88} and \citet{n88} at high metallicity is discussed in relation to two extreme open clusters, NGC 6791 and NGC 6253. It is undoubtedly the case that with the availability of VF, new and totally valid attempts will be made to generate revised, multi-term functions for metallicity as a function of $b-y, m_1$, and $c_1$ following the approach initiated by \citet{sn89} and built upon by MA and N04. The view of this investigation is that the data have evolved to a point where the efficacy of having one (or two) function(s) that fit all stars at all temperatures, surface gravities, and abundances has been reduced. Recreating the complex interdependence of the parameters leads to the necessity of adding additional terms that can, in some regions of the parameter space, hide real effects while creating unnecessary errors in others. If the Str\"{o}mgren system remains a primary survey technique for identifying stars of different classes of interest based upon fundamental parameters with a potential for future followup with larger and more competitive telescopes, it is important that such classifications remain as clearcut and definitive as possible, minimizing the number of false positives and negatives.

The intent of this paper is to investigate the reality of the cool dwarf/giant confusion and to determine if a photometric means can be constructed that will eliminate the degeneracy while enhancing the metallicity sensitivity of the $uvby$ system. With this in mind, we will use Sec. 2 to present the compilation of the spectroscopic and photometric database for the analysis and Sec. 3 to define the extent of the problem. Sec. 4 presents a potential means for addressing the problem and Sec. 5 summarizes our conclusions.

\section{CALIBRATION DATA}
A preliminary discussion of the photometric and spectroscopic samples may be found in \citet{att07}. Because of additions and revisions, the compiled data presented below supercede the earlier work.

\subsection{$uvby$H$\beta$ Photometry}
To optimize the reliability of any calibration, one needs to include large samples of data with high internal precision. While those two characteristics are often difficult to come by simultaneously, on the $uvby$H$\beta$ photometry front the dedicated work of a number of observers over the last 25 years has produced an ideal data ensemble. The defining core catalogs are those of \citet{o83,o93,o94a,o94b}, with the redder stars ($b-y$ $>$ 0.45) of the first source transformed to the system of the other three. To this core set has been added the data from \citet{cb66,go76,go77,tw80,op84,sn88,pe91,sc93}, each catalog transformed to the core catalog system using the direct comparisons derived in the literature. Multiple observations of the same star from different catalogs were then averaged using a weighting scheme based upon the inverse of the standard error of the mean for each index. The final composite catalog contains just over 35000 stars. While a number of additional $uvby$H$\beta$ catalogs were tested, they were ultimately excluded because the derived scatter in the residuals between the data and the core set was found to be significantly larger than calculated for the list above.  

Before discussing the spectroscopic data, a few items should be mentioned regarding \citet{tw80}. Since the original publication of photometry for 1007 stars, comparison with \citet{o83} and a comparison of the entire catalog to the published data on its stars has generated a modest list of changes, reducing the original sample to 995 stars. The star listed as HD 1326 is HD 13246; its photometry has been averaged with the data for that star. The star listed as HD 8224 is HD 8178; its photometry has been averaged. Star HD 25059 is actually HD 25054. HD 31265 is HD 32065; its photometry has been averaged with the data for that star. HD 85124 is HD 85424; the photometry has been averaged. The two data sets listed for HD 206481 have been averaged. HD 221551 is HD 221552; the data have been averaged. HD 182028 is HD 183028; the data have been averaged. Stars dropped as misidentifications include HD 1963, HD 101085, HD 108500, HD 167854 and HD 218482.

\subsection{The Spectroscopic Catalog}
The success of any statistical analysis of a large data sample generated by the merger of data from multiple sources, particularly one tied to the definition of a metallicity scale, is invariably driven by the degree of homogeneity that can be imposed upon the merged components. Approaches to such mergers for abundance catalogs are varied, ranging from the very basic \citep{ca01} to the more elaborate \citep{bt}. Our approach lies somewhere in the middle and represents a modified version of the technique used in the recalibration of the DDO system \citep{tw96}. The critical component in our approach is the existence of a large base catalog of homogeneous high dispersion spectroscopic abundances obtained by one group using the same reduction technique for all stars. For the red giant discussion, the data of \citet{mw90} proved ideal; for the dwarfs, the exquisite sample of VF has proven invaluable. Given the base catalog, the next step is the transformation of additional sources to the base standard. For the giants, issues of reddening were important due to the range in distance among the sample, the impact upon the temperature scale, and ultimately the abundance. For the dwarfs, virtually all of the stars are within 100 pc and, on average, reddening effects should be small to nonexistent. For the sake of simplicity, we assume that whatever corrections have been made for reddening within the individual sources, even if they contradict each other, are correct and merely add to the scatter among the comparisons. In contrast with the giants, however, with only a few exceptions, all the stars in the dwarf catalog have been observed by $Hipparcos$ \citep{pe97} and therefore have some form of distance and absolute magnitude estimate. These have been used in some cases to generate surface gravity estimates used in analysis of the spectra, while others have used the internal consistency of the line analysis to define a log g. In deriving our transformations to the base catalog, we have therefore attempted to optimize the match by using the following relation:

\medskip
\centerline{[Fe/H]$_{VF}$ = $a\ $[Fe/H]$_{ref}$ + $b\ $log$\ $T$_e$ + $c\ $log$\ $g + $d$}
\medskip

In some cases, the significance of the surface gravity and/or effective temperature terms was negligible and they were dropped from the final transformations. It's encouraging that the linear slopes were normally close to 1.0 and the offsets consistently below 0.10 dex.

To create the catalog, the literature in recent years was surveyed and papers with high-dispersion spectroscopic abundances for large samples of dwarfs were cross-correlated with VF. If the source catalog exhibited adequate overlap with VF and, after applying a transformation equation of the form noted above, produced a scatter among the residuals between the source catalog and VF below 0.065 dex, the catalog was retained. Of 35 spectroscopic surveys tested, 25 were retained. Of the remaining surveys, five were excluded because they focused primarily on metal-deficient halo dwarfs and had too little overlap with VF to define a reliable transformation, while five had dispersions among the residuals in the comparison that were deemed too large. 

Adopting an approximate estimate of $\pm$0.025 dex as the typical uncertainty within the catalog of VF, the error in the residuals was used to define an approximate internal error for each catalog. The data for each catalog were then transformed to the system of VF and the abundances for each star averaged using a weighting scheme based upon the inverse of the adopted internal errors for each catalog. Each survey was then transformed to a new composite catalog created by including all surveys except the one undergoing the transformation. The newly transformed 25 spectroscopic samples and VF were then merged a final time to create the spectroscopic abundance catalog of 1801 stars. Table 1 contains a summary of the information about each survey merged into the final catalog, including the number of stars in the transformation, the number of abundances contributed to the final catalog, the calculated internal errors associated with the abundances from the catalog, and the transformation coefficients as defined above. Because 1039 of the stars are found in the core catalog of VF, the majority of the final abundances should have uncertainties below 0.03 dex and all should be below 0.06 dex. 

While not a key focus of this analysis, for purposes of the discussion of the sensitivity of the various photometric indices, the adopted effective temperatures and surface gravities for each star from each survey were also transformed to the system of VF using only linear relations in $T_e$ and log g.  

The final catalog of homogeneous abundances was then matched with the composite catalog of $uvby$H$\beta$ data. Because H$\beta$ photometry has been published for only a portion of the stars in the photometric catalogs, an additional check for this important data was made by doing a match with the database compiled by N04.  Of the 1801 stars in the abundance catalog, 1587 have $uvby$ indices from the sources noted above. Of these, H$\beta$ photometry is available for 1298 stars. Note that this is a 50$\%$ increase over the preliminary sample discussed in \citet{att07}. The distribution of stars as a function of [Fe/H] and $b-y$ are shown in Figs. 1 and 2, respectively. In Fig. 1, the three histograms illustrate the distribution for all stars, stars with $uvby$ photometry, and stars with $uvby$H$\beta$. For Fig. 2, the two histograms are stars with $uvby$ and stars with $uvby$H$\beta$.  Two features stand out. First, the number of stars with [Fe/H] below $-$1.0 is too small to be significant. Therefore, our conclusions regarding the metallicity sensitivity of the indices will only apply to traditional disk stars, i.e., [Fe/H] above $-1.0$. Even with this restriction, the reliability of any statements will decline progressively as one refers to stars with metallicity below $-0.50$. Second, the fraction of stars with H$\beta$ photometry drops dramatically for $b-y$ redder than 0.44. As we will show in Sec. 4, this deficiency is unfortunate in that it represents a missed opportunity to improve the applicability of the Str\"{o}mgren system for metal-rich dwarfs at supposedly cool temperatures.

\section{THE PROBLEM: DERIVING [Fe/H] FOR COOL STARS}
\subsection{Current Calibrations: Global Properties}

As discussed in the Introduction, the evolution of the $uvby$ system has generated two dominant current functional forms for converting the observed indices to metal abundance, those of MA and N04. To begin the delineation of the problems addressed by this investigation, we first do a simple comparison of the predicted abundances from the two photometric functions with the spectroscopic sample for all stars with $uvby$ data. No reddening corrections will be applied since, for the cool stars where the problems arise, the vast majority of the sample lies within 100 pc and reddening should be small to negligible. This also avoids potential issues with reddening determination from intrinsic colors, a point we shall return to in Sec. 4. For N04, calibrations are supplied for F stars ($b-y$ $<$ 0.38) and G stars ($b-y$ $>$ 0.44). To bridge the gap between these two, the abundance has been derived using both functions, with the two values averaged using a simple weight based upon the fractional position in $b-y$ between the end points of the calibrations.

Fig. 3 shows the trend for the average residual in [Fe/H], in the sense (SPEC-PHOT), as a function of $b-y$. Filled circles are the data for MA and filled triangles are the data for N04. To highlight the differences in the two functions, we also plot the dispersion among the residuals at each color as open symbols for each function. Finally, the dispersion in the spectroscopic [Fe/H] sample at each color is plotted as a star; this effectively tells us what the scatter would be if we had a photometric calibration that produced the same metallicity for every star.

A number of trends are apparent:

a) The photometric abundances of  MA (filled circles) typically underestimate [Fe/H] by about 0.08 dex, with a very weak trend of a decreasing differential with increasing color. Such an offset is not unexpected given that the original spectroscopic catalog used to define the calibration \citep{ca01} and that of VF have independently defined zero-points. The exception to the trend is in the last bin at the cool end containing 22 stars; the size of the sample is such that one or two deviant points can easily distort the mean and the dispersion, coupled to the fact that the calibrations are least well determined for this color range. One star in this bin, HD 38114, has been excluded from the comparisons because the residual difference between the spectroscopic and photometric abundance is almost 1.0 dex. Its photometric indices are indicative of an unevolved cool dwarf with [Fe/H] near $-1.0$, while it has been spectroscopically analyzed as subgiant with [Fe/H] = 0.0. It deserves a closer look spectroscopically.
 
b) The photometric abundances of N04 (filled triangles) also underestimate the metallicity at a comparable level, but the trend with color is significantly more structured. In particular, between $b-y$ = 0.35 and 0.39, the offset increases by 50$\%$ to 0.14 dex. A similar jump occurs among the coolest stars ($b-y$ $>$ 0.53) where the offset grows from 0.07 to 0.11 dex. 

c) The dispersion in the abundance residuals using MA (open circles) is almost constant between 0.07 and 0.10 dex for $b-y$ bluer than 0.43. There is sharp increase in the dispersion among the residuals to 0.12 dex that sets in for $b-y$ redder than 0.43 and grows slowly with increasing color. For the hotter stars, the dispersion predicted from photometric scatter alone is between 0.05 and 0.07 dex, so the observed spread among the residuals confirms both the small uncertainty in the composite spectroscopic abundances and the exceptional capability of $uvby$ data to supply reliable abundances for F dwarfs to the limit of the photometric accuracy of the catalog. 

d) The dispersion in residuals using N04 (open triangles) follows a pattern similar to that for MA, with the distinction that the jump in the dispersion kicks in at a bluer color, i.e., near $b-y$ = 0.35. Beyond $b-y$ = 0.43, the two calibrations have very similar general characteristics. However, all things being equal, the MA calibration is to be preferred over the full range of color, requiring only a constant offset to bring the photometric abundances onto the same scale as VF.

As noted above, the photometric metallicity determinations are approaching their limits in accuracy for the hotter stars, but it would be valuable if one could decrease the dispersion among the stars cooler than the sun. Moreover, near the cool end of the sample, the growth in the dispersion among the residuals approaches the dispersion in the spectroscopic [Fe/H] among the stars in the sample, indicating that the calibrations convey little more information than would be implied if the stars in the bin all had the same metallicity. Finally, while the global properties of the calibrations appear to be satisfactory, the question remains as to whether or not residuals exhibit a dependence on [Fe/H]; in other words, is the scatter primarily caused by stars at the extreme ends of the metallicity scale where the numbers are invariably small?  As discussed in Sec. 1, it was the absence of a significant sample of metal-rich stars at all $b-y$ that generated the initial problem \citep{tw02} with the global solution of \citet{sn89}.  

To investigate this possibility, the sample was sorted by $b-y$ and the residuals, in the sense (SPEC-PHOT) plotted as a function of [Fe/H]. For $b-y$ $<$ 0.39, no trend was found beyond the offsets noted in Fig. 3. However, for $b-y$ between 0.39 and 0.50, a clear pattern emerged, as shown in Fig. 4. For stars in this color range, the metallicity calibration of MA compresses the metallicity scale and therefore the metallicity distribution; stars with [Fe/H] near $-1.0$ are made more metal-rich by approximately 0.1 dex, while stars with spectroscopic abundances near +0.40 are found to have [Fe/H] $\sim$ +0.25. The exact correction to the photometrically derived abundances is 

\medskip
\centerline{[Fe/H]$_{SP}$ = 1.18$\ $[Fe/H]$_{PH}$ + 0.11}
\medskip

We will discuss the trend for the reddest stars below.

\subsection{The $m_1-c_1-[Fe/H]$ Relations}
To understand the nature of the problems among the cooler stars, we will subdivide the sample into three generic groups called hot ($b-y$ $<$ 0.43), warm (0.43 $\leq$ $b-y$ $<$ 0.50) , and cold ($b-y$ $\geq$ 0.50), and illustrate the role that $m_1$ and $c_1$ play in defining the photometric metallicity for each group. Because there is a non-negligible color dependence on the sensitivity, we will use a subsample within each category to represent the full range: $b-y$ = 0.30 to 0.329 for hot, 0.44 to 0.469 for warm, and 0.51 to 0.549 for cool.

Fig. 5 plots the trend of spectroscopic abundance with $m_1$ for the hot group, while Fig. 6 shows the variation of [Fe/H] with $c_1$. The inherent value of the $m_1$ index as a measure of metallicity is apparent. There is no significant variation of the scatter with metallicity over the entire range from [Fe/H] = $-0.8$ to +0.3, with every expectation that extrapolation to [Fe/H] = +0.4 to +0.5 should pose no problems. The corresponding $c_1$ relation shows a modest trend with [Fe/H], in part tied to the correlation between $m_1$ and $c_1$ across the color range. For the 120 stars in this bin, the residuals between the spectroscopic data and the photometric calibration of MA, in the sense (SPEC-MA), average +0.090 $\pm$ 0.089. Surprisingly, this can be improved upon through the use of a simple linear function. With [Fe/H] = $-2.480$ + 10.15$m_1$ + 1.972$c_1$, the comparable residuals average 0.000 $\pm$ 0.077. No cross terms of significance were found. 

Turning to the warm stars, Figs. 7 and 8 illustrate the comparable trends for [Fe/H] as a function of $m_1$ and $c_1$. The reversal compared to the hot stars is dramatic. The metallicity dependence is now dominated by $c_1$, while $m_1$ exhibits, at best, a modest correlation with [Fe/H], with a large scatter at a given $m_1$, confirming the analysis of a smaller sample in \citet{tw02}. The $c_1$ behavior was expected based on the observations of metal-deficient dwarfs by \citet{sn88} who showed that at a given $b-y$, more metal-poor stars on the main sequence have lower $c_1$ values; this trend evidently extends over the entire range in [Fe/H]. A basic explanation is tied in part to the metallicity dependence of $b-y$. The more metal-rich stars at a given $b-y$ are intrinsically hotter, on average, higher on the main sequence, and brighter. Since $c_1$ is well-correlated with luminosity, all things being equal, larger [Fe/H] implies higher $c_1$. As we will discuss in the next section, model atmospheres predict that at a given temperature, $c_1$ will increase with metallicity before reaching a maximum and declining at even higher [Fe/H]. The maximum $c_1$ and the start of the decline occurs at a lower [Fe/H] for cooler temperatures. This enhances the apparent $c_1$ correlation with [Fe/H]. 

The disappointing aspect of the transition from hot to warm stars is the apparent weak, though not absent, metallicity sensitivity to changes in $m_1$. A calibration of [Fe/H] using quadratics in $c_1$ and $m_1$, [Fe/H] = $-$6.06 + 15.72 $c_1$ +18.57 $m_1- 16.40\ c_1$$^2$ - $31.07\ m_1$$^2$, applied to 148 warm stars has a mean among the residuals of 0.000 $\pm$ 0.098 dex; the same sample processed using the calibration of MA has a mean of +0.078 $\pm$ 0.111. Inclusion of linear or quadratic cross terms alters the dispersion in the residuals by less than 1$\%$.

The color range to test the cool stars is slightly larger and the sample size is smaller, 49 stars. The complete set of points in Fig. 9 implies no [Fe/H] sensitivity to changes in $m_1$; stars with $m_1$ between 0.3 and 0.4 can be any metallicity between [Fe/H] = $-0.7$ and +0.5. In contrast, following the pattern set up by the warm stars, there is some correlation in Fig. 10 between [Fe/H] and $c_1$, with more metal-rich stars having, on average, higher $c_1$. However, compared with Fig. 8, the scatter is significantly larger at a given [Fe/H]. 

A partial explanation of the source of the scatter is apparent if one isolates unevolved dwarfs from subgiants using their absolute magnitudes as derived from their $Hipparcos$ \citep{pe97} parallaxes. Because of the redder colors, the stars plotted now include stars on the unevolved main sequence (filled circles) with $M_V$ $\sim$ 6 alongside subgiants and stars near the base of the giant branch (open circles) with $M_V$ $\sim$ 3. From Figs. 9 and 10, the data points now crudely separate into two distinct groups. In $c_1$, the trend with [Fe/H] for the dwarfs bears a strong resemblance to the curve for the warm stars. For $m_1$, there is an approximately linear relation among the dwarfs for [Fe/H] = $-0.7$ to +0.3. However, in both figures, for dwarf stars above [Fe/H] = +0.3, the indices resemble those expected from an extrapolation of the pattern for the evolved stars. This sharp distinction is best seen in the $m_1$ plot, where the very metal-rich dwarf stars lie approximately 0.2 mag lower in $m_1$ than one would expect for an extrapolation of the linear dwarf relation.  In short, very metal-rich dwarfs occupy the same region of the $m_1$, $c_1$ diagram as metal-rich subgiants, confirming the trend found by \citet{tw02}.

To emphasize this point, we reconstruct two plots from \citet{tw02} with the much larger and more accurate current sample of photometry and abundances. Fig. 11 is the $c_1, b-y$ distribution for all stars with [Fe/H] $\geq$ 0.00. The lower limit was chosen simply to minimize the confusion of points in the figure and does not affect the conclusions. Dwarfs with [Fe/H] $\geq$ +0.25 are plotted as filled circles while subgiants in the same metallicity range are open circles.  All stars with [Fe/H] $<$ +0.25 are crosses.

The discontinuity in the main sequence distribution is obvious. Eight very metal-rich unevolved dwarfs between $b-y$ = 0.50 and 0.58 lie in a region normally populated by subgiants and giants, as if a piece of the main sequence has been shifted diagonally toward higher $c_1$ and redder color. Only one very metal-rich star populates  the unevolved main sequence as defined by the stars with [Fe/H] between 0.0 and +0.24 between $b-y$ = 0.50 and 56; it has a spectrocscopic abundance of [Fe/H] = +0.25. It should be mentioned that one additional very metal-rich star, HD 17006, also lies in the subgiant region, but the photometry for this star does not come from one of the primary photometric sources and is not included in our catalog, so it has not been included in the plot. Equally striking is the return to a normal separation of subgiants and dwarfs for $b-y$ redder than 0.60. Note that $every$ star with [Fe/H] $<$ +0.25 with $b-y$ redder than 0.50 and $c_1$ $>$ 0.38 is an evolved star, i.e., a subgiant or a giant.

Fig. 12 shows the analogous plot for $m_1, b-y$. The pattern is virtually identical, though a second star in the critical color range has an $m_1$ index closer to the dwarf sequence than the subgiants. The very metal-rich subgiants follow the extension of the metal-rich subgiants to redder $b-y$ and higher $m_1$, but from the $m_1$ diagram the unevolved dwarfs  would be classed as either metal-poor dwarfs or very metal-rich subgiants. 

How does the calibration of MA handle this discontinuity? Fig. 13 shows the residual plot for stars between $b-y$ = 0.50 and 0.57, the limit of the MA calibration. The trend is an excellent reflection of the pattern implied by Fig. 10. While the scatter is significant, the unevolved dwarfs (circles) generally scatter around a mean residual near 0.0 with no obvious separation of the dwarfs with [Fe/H] $>$ +0.25 (filled circles). The evolved stars (crosses) are offset toward positive residuals by about 0.15 dex, implying that the MA calibration will systematically underestimate [Fe/H] for cool subgiants by $\sim$0.15 dex relative to dwarfs at the same color.

In summary, the key to successful application of the MA calibration (or any calibration tied solely to $uvby$ indices) for cool stars appears to be the separation of unevolved dwarfs from subgiants and giants. The ability to separate cool stars into luminosity classes has long been one of the strengths of the $uvby$ system, but is also a requirement for basic transformation of instrumental indices to the standard system for photoelectric photometry \citep{o93,tat95} or CCD data \citep{at00a,at00b}. To determine the simplest option for sorting the cool stars by luminosity class, the 142 stars redder than $b-y$ = 0.50 were sorted by log g and various combinations of $c_1$ and $m_1$ indices tested, with the knowledge that generally $c_1$ is larger and $m_1$ is smaller for evolved stars at a given [Fe/H]. Using log g = 4.2 as the breakpoint for a dwarf versus a subgiant, the luminosity class parameter, LC, defined as

\medskip
\centerline{LC $= c_1 - 2.0\ m_1 + 3.0\ (b-y) - 0.15$}
\medskip

identifies all dwarfs between $b-y$ = 0.50 and 0.70 as stars with LC $<$ 1.0.  Fig. 14 shows the plot of LC for the cool star sample. Filled circles are stars with log g $>$ 4.20 while crosses are stars with lower surface gravity. Note, there is one cross at $b-y$ = 0.57 with LC $<$ 1.0; this star, HD 45088, has log g = 4.0, but is a dwarf based upon its absolute magnitude and $c_1$. It appears that the LC parameter has correctly identified this star as a dwarf. There are 9 stars with log g indicative of unevolved dwarfs that fall within the giant range. One star at $b-y$ = 0.50, HD 182736, has log g = 4.27, just over the border for classification. Again, its absolute magnitude confirms that the LC classification is correct and the star is, in fact, a subgiant. The remaining 8 stars are the super-metal rich dwarfs with [Fe/H] = +0.25 or larger, confirming again the difficulty of isolating this extreme group of stars from the typical field giant at the same $b-y$. There is a tendency for the metal-rich dwarfs to lie near the base of the giant distribution, but this fails for the dwarfs with $b-y$ below 0.53 because the $c_1,b-y$ relation for the metal-rich dwarfs crosses that for the giants in this color range, removing the differential in $c_1$ found among the cooler stars as shown in Fig. 11. One can improve matters slightly by selecting only stars with $c_1$ $>$ 0.36 first, then plotting $m_1$ versus $b-y$, as in Fig. 12. On average, the metal-rich dwarfs have slightly higher $m_1$ than the giants, though the separation is disappointingly small.

 \section{ENHANCING THE METALLICITY CALIBRATION}

The implication of the previous section is that the $uvby$ system offers a great deal by allowing one to determine metallicity and luminosity for cool dwarfs and subgiants, but the full potential of the system is hampered, in part, by the changing interdependence of the three primary indices as one moves down the main sequence and the discontinuous impact of increased metallicity on the indices of cool dwarfs. Universal functions of the type proposed by MA and N04 can supply reasonable abundance estimates, but the need to smooth over the rapid changes in sensitivity with $b-y$ can weaken the value of the estimates at some colors and wash out sharp discontinuities of the type identified for very metal-rich dwarfs. As stated earlier, part of the transition in sensitivity between $m_1$ and $c_1$ arises from the metallicity impact on $b-y$. Since $m_1$ increases with $b-y$ along the main sequence but $c_1$ decreases compared to standard relations at a fixed [Fe/H], shifting $b-y$ to the red due to increased metallicity alone will make the $c_1$ differential compared to the standard relation increase  while that for $m_1$ appears more metal-poor.  A subtle but important correlation is that the coupled changes in $b-y$, $m_1$, and $c_1$ are qualitatively similar to those caused by reddening. While the net impact on the latter pair of indices depends on the direct effect of varying the metallicity on the $c_1$ and $m_1$ indices, changes that can vary in size and direction as a function of temperature and [Fe/H] independent of the changes in $b-y$,  it may improve the sensitivity of the system if one removes the metallicity-dependent color shift in $b-y$.

A common solution for F stars that enhances the metallicity sensitivity while decoupling the $m_1$ and $c_1$ indices from the $b-y$ temperature dependence has been the adoption of H$\beta$ as the primary temperature index. H$\beta$ also has the advantage of being reddening and, supposedly, metallicity independent. Though it has been extensively used for hotter dwarfs, it has seen less application for G and K dwarfs cooler than the sun, no doubt a product of the belief that the Balmer lines of H should be weak in cooler stars, rendering the index ineffective. Empirically, from a sample of 289 stars with independently derived temperatures, \citet{al96} found this not to be the case. Over the entire range of interest, 4000 K to 7000 K, the H$\beta$ indices maintained an almost constant sensitivity to changes in temperature, with only a weak metallicity effect amounting to a shift of 0.015 to 0.020 mag/dex in H$\beta$. It therefore seems plausible that the successful approach of adopting H$\beta$ for hotter stars should carry over into the cool dwarf regime and minimize, if not remove, part of the sensitivity reversal exhibited in Figs. 5 through 10. 

To test this assumption, we approached the problem from two directions, deriving  indices from synthetic spectra covering a wide range in temperature and metallicity, and empirical tests using the observed indices for stars of varying temperature and spectroscopic metallicity. The former approach is crucial because the dramatic decline in the number of cooler dwarfs with H$\beta$ indices as seen in Fig. 3 makes testing the full range of parameters difficult.

\subsection{Synthetic Indices}
To gauge how much of the derived variation of the photometric parameters is a product of the spectrum synthesis programs rather than an indication of real stellar changes, we used two high resolution synthetic spectral libraries, one created with the PHOENIX (PH) spectral synthesis code \citep{bh05} and the second based upon MARCS (MC) (see, e.g. \citet{gu03}), to generate synthetic $uvby$H$\beta$ colors and indices for K through F dwarfs (4500 K $<$ T $<$ 7000 K, log g=4.5) at metallicities of [Fe/H] = $-0.5$ and higher. In light of the dominance of the thin disk population of stars in the solar neighborhood, all models evaluated had scaled-solar elemental distributions. For more information about the spectral libraries, we refer the reader to \citet{be05}. 

In order to minimize systematic effects in the calculation of synthetic spectra, the $uvby$ transmission functions from \citet{cb70}, as well as the wide and narrow H$\beta$ transmission functions for KPNO filters numbered 1568 and 1567, respectively, were modeled with either polynomials or gaussians, so that each spectral flux point was convolved with an accurate transmission factor. Using the spectral flux, the synthetic magnitudes through each filter are calculated with 

\medskip
$m_{\lambda}=-2.5\ log\sum_{\lambda}f(\lambda)S(\lambda)\Delta{\lambda}$
\medskip

where $f(\lambda)$ refers to the spectral fluxes and $S(\lambda)$ the transmission functions for each filter, respectively, with wavelength. The wavelength resolution of the synthetic spectra is high enough, 0.2 nm or less for both sets, that the simple numerical integration adopted above at each wavelength point is more than adequate to produce accurate magnitudes with minimal error. The synthetic colors are then defined using appropriate combinations of the calculated magnitudes.

While we are primarily interested in the variation of the indices with changing stellar parameters, it is useful to match the synthetic colors as closely as possible to the absolute photometric system. To accomplish this, the synthetic indices were treated in much the same way one would treat instrumental photometry, a reliable match to the standard system but requiring some transformation relations to account for subtle differences between the observations and the standards.
The transformation from the synthetic instrumental system to the standard system was performed as follows. For $b-y$, $v-b$, and $u-v$, linear color-temperature relations were derived using stars with \citet{o83,o93,o94a,o94b,no04} 
$uvby$H$\beta$ photometry and VF spectroscopic metallicities and temperatures. The selected stars range in temperature from 4500 K to 7000 K,  log g $\geq$4.3 to remove evolved stars, and  $-0.1 \leq$  [Fe/H]$ \leq +0.1$. 
These color-temperature relations were then used to construct transformation equations between the synthetic indices and the standard system using the assumed temperature of the synthetic spectra at [Fe/H] = 0.0 as the common link. Although the transformations were essentially linear, we chose to retain higher order polynomial terms, since the deviations from linearity were small but noticeable. These instrumental transformation equations were then applied to the synthetic colors at all metallicities and final values of $m_1$ and $c_1$ derived at all $b-y$ and [Fe/H]. 

The transformation procedure for $H\beta$ was similar, but with two differences. 
First, we used the H$\beta$-temperature relation from \citet{al96} due to the paucity of $H\beta$ data for stars cooler than T $\sim$ 5000 K in the photometric catalogs. Second, we calculated transformations between synthetic and standard $H\beta$ indices at $both$ [Fe/H] = $-1.0$ and [Fe/H] = 0.0. When converting synthetic H$\beta$ to standard H$\beta$ indices at a given metallicity, we computed the standard values using both transformations, then interpolated or extrapolated to the expected value based linearly on the metallicity of the theoretical spectrum. The only case of extrapolation arises when obtaining 
standard H$\beta$ indices at [Fe/H] $>$ 0.0. Although imperfect, it appears this is 
necessary due to the highly non-linear transformation between the synthetic and 
standard H$\beta$ indices. In the discussion that follows, we will illustrate the results using the MC spectra, but note any differences with the indices generated from the PH compilation.  

The use of transformation calibrations between the synthetic and standard systems implies that the synthetic colors and indices at [Fe/H] = 0.0 are a  match to real stars. While this claim has variable validity depending upon the wavelength region, stellar properties, and the level of accuracy desired, it should still allow us to get a reliable handle on the qualitative $relative$ effects on the indices that may appear as the metallicity varies. As a first test, we look at the dependence of $c_1$ with varying [Fe/H] at a fixed value of H$\beta$ in Fig. 15. Symbols represent the synthetic indices at fixed H$\beta$ with open circles, stars, open squares, filled triangles, and crosses defining H$\beta$ = 2.56, 2.58, 2.60, 2.62 and 2.64, respectively. It should be kept in mind while reviewing the trends that as H$\beta$ increases, the actual data point scatter about the mean curves grows as luminosity effects potentially begin to play a dominant role among the hotter stars. 
Moreover, because $c_1$ includes the $u$ filter, deviations between theory and observation are not unexpected.
  
The models predict that above [Fe/H] = $-0.5$, $c_1$ should decrease as H$\beta$ decreases (temperature drops). More important, as [Fe/H] rises from $-0.5$, $c_1$ is predicted to remain constant or rise slowly to a maximum before declining at [Fe/H] above solar. The trend with [Fe/H] is modest enough that the expectation is that $c_1$ will have at best only a weak dependence on [Fe/H] for stars in the disk and, if anything, should decline for super-metal-rich stars at cooler temperatures. The reality is shown via the solid curves derived by drawing mean relations through the actual data contained in bins 0.01 mag wide centered on H$\beta$ = 2.56 (solid), 2.58 (dotted), 2.60 (dashed), and 2.62 (solid) (bottom to top). Also shown as filled circles is the complete sample for H$\beta$ = 2.535 to 2.549. Empirically, the qualitative trend  of $c_1$ with [Fe/H] holds up in that, on average, at a given [Fe/H], $c_1$ does increase as stars get hotter. Moreover, the trend of $c_1$ with [Fe/H] remains modest to nonexistent at a given value of H$\beta$ for all but the coolest stars in the sample. In contrast with the theoretical predictions, the observed trends diverge toward lower [Fe/H], though for the cooler stars, there are too few stars with [Fe/H] above +0.3 and H$\beta$ photometry to reliably predict the pattern for super-metal-rich dwarfs. The data plotted for the coolest stars again exhibits a rise with [Fe/H] that is virtually identical to that for the H$\beta$ = 2.56 sample, with some indication that $c_1$ hits a maximum before declining. In summary, both theory and observation indicate that switching to H$\beta$ as the primary temperature index significantly reduces the metallicity sensitivity of $c_1$ at all temperatures, though higher [Fe/H] generally implies higher $c_1$ at a given temperature and surface gravity.

Turning to $b-y$ in Fig. 16, the agreement between theory and observation is much better. The symbols are the same as in Fig. 15, though the hotter stars (larger H$\beta$) are at the bottom and temperature decreases vertically. Surprisingly, the data show that the correlation between $b-y$ and [Fe/H] at a given H$\beta$ has virtually the same shape at all H$\beta$; what changes for each curve is the zero-point, with $b-y$ increasing as temperature declines. The change in $b-y$ with [Fe/H] is modest between [Fe/H] = $-$1.0 and $-$0.1, with a sharp rise at higher [Fe/H]. These results confirm the preliminary trends discussed in \citet{att07} and the data for the coolest stars demonstrate that the pattern continues for H$\beta$ below 2.55. There is a modest offset between the observed and theoretical curves and the theory predicts that the relations should show decreasing separation as H$\beta$ declines, opposite to what is observed. It should be noted, however, that the latter discrepancy depends upon the choice of synthetic spectra. While the $b-y$ indices based upon PH spectra exhibit almost identical morphology with [Fe/H] at all H$\beta$, the curves show increasing separation at lower H$\beta$.

An important conclusion from this analysis is illustrated by the thick, solid straight line in Fig. 16. The line is the predicted trend of $b-y$ with [Fe/H] for stars with H$\beta$ = 2.60 using the average of the intrinsic color relations of \citet{o88} and \citet{n88}. As discovered in \citet{att07}, the traditional color relations do well at predicting the intrinsic colors between $b-y$ = $-$1.0 and $-$0.1, but fail to account for the changing slope at higher [Fe/H], with the result that more metal-rich dwarfs are increasingly predicted to be bluer than they are, leading to a growing overestimate of the reddening value to explain the discrepant colors, as was found for NGC 6791 \citep{att07} and NGC 6253 \citep{tw03}. Near [Fe/H] = +0.45, the discrepancy is close to 0.07 mag in $b-y$, generating a reddening error of 0.10 mag in E$(B-V)$. 

We close by noting that almost all super-metal-rich stars ([Fe/H] $\geq$ 0.25) with $b-y$ between 0.50 and 0.58 are predicted to have H$\beta$ between 2.56 and 2.60, well within the commonly applicable range of the index, and turn to the metallicity calibration as defined using H$\beta$ as the temperature-dependent variable.

\subsection{The H$\beta$-Based Metallicity Calibration}
The photometric sample was sorted by H$\beta$ in bins 0.011 mag wide centered at H$\beta$ = 2.635,2.625, 2.615, etc. down to 2.555. The bin width was selected to allow some overlap for each sample with the contiguous bins. Using the same approach outlined in MA to calibrate [Fe/H], a polynomial function in $b-y,m_1$, and $c_1$ with terms as high as cubics in each variable, along with potential cross terms, was tested for each bin. As expected, the majority of terms were discarded as inconsequential, defined as changing the dispersion in the residuals by less than 2$\%$ if included in the function. The largest number of terms was five, defined as follows

\medskip
\centerline{[Fe/H] = $a$ +  $b\ m_1$ + $c\ m_1$$^2$ + $d\ c_1$ + $e\ (b-y)$}
\medskip

Even more important, as hoped, the value of $d$, the coefficient on the $c_1$ term, remained small to modest across the entire H$\beta$ range. The dominant term in the metallicity calibration remains $m_1$, though the impact of $c_1$ is real, particularly at the cooler end. The coefficients at each 
mean H$\beta$ are listed in Table 2, along with the number of stars included in each bin and the dispersion in the residuals in [Fe/H] between the spectroscopic and photometric values. In defining the calibrations, two additional stars with anomalously large residuals, HD 35850 and HD 110898 were excluded; the spectroscopic abundance and temperature for HD 35850 are too high for its photometric indices, while HD 110898 is too metal-poor. The dispersion in the residuals ranges between 0.061 dex and 0.085 dex over the entire range from H$\beta$ = 2.64 to 2.55. The small dispersion among the residuals is encouraging, though it should be noted that this stellar sample is about 25\% smaller than the one used for the $b-y$ analysis and many of the more extreme stars lack H$\beta$ estimates. Moreover, switching to H$\beta$ does not diminish the potential for systematic errors with [Fe/H] of the type illustrated in Fig. 4. As a test, we have plotted the residuals as a function of [Fe/H] with the stars sorted by H$\beta$ and find no statistically meaningful trend for stars based upon their location in the temperature range. Keeping in mind the small sample of stars in this range, there is modest evidence that at the metal-rich end ([Fe/H] $\geq$ +0.10), the photometric abundances should be adjusted using the relation

\medskip
\centerline{[Fe/H]$_{SP}$ = 1.61$\ $[Fe/H]$_{PH}$ $-$ 0.06}
\medskip

implying that without the correction, the photometric calibration will systematically underestimate the abundances of super-metal-rich stars by approximately 0.10 to 0.15 dex.

While the H$\beta$-based calibration supplies abundances with encouragingly small residuals, it also has one additional benefit that is often overlooked in the discussion of polynomial fits, the sensitivity to systematic errors, in particular, the impact of reddening on the final abundance. We have assumed throughout this discussion that reddening for the stars involved is small to negligible. This is almost certainly true for the nearest stars, but what if observations are made of stars at greater distance and reddening corrections are ignored or inadequately applied? The impact of underestimating E$(B-V)$ by 0.020 mag (E$(b-y)$ by 0.015) is shown in Fig. 17, where we have plotted the residuals between the true photometric [Fe/H] and the calculated photometric [Fe/H] with an underestimate for the reddening for the calibration of MA (crosses) and the calibrations of Table 2 (open circles). For the hotter portion of the sample (H$\beta$ $>$ 2.60), the error in the calculated [Fe/H] is typically 3 times larger for MA. For the cooler side of the sample, there is significant structure for the H$\beta$-based calibration, but the MA offset remains, on average, 20$\%$ to 200$\%$ larger. Thus, use of the MA calibration on stars with random errors of $\pm$0.02 in E$(B-V)$ will add unnecessary scatter to the final abundance distribution, over and above the effect caused by random photometric errors.

\section{SUMMARY AND CONCLUSIONS}
The reliability of any photometric system in generating fundamental stellar properties is invariably limited by the photometric accuracy of the indices. Under the assumption that one can obtain $uvby$H$\beta$ data with uncertainties typically below 0.010 mag in every index, the next important factor is the effectiveness in disentangling the interplay of the stellar properties in each index, in the hope of isolating each parameter with the greatest degree of sensitivity. Over the last 40 years, the evolution of the $uvby$H$\beta$ system has been driven (and restricted) by the expansion of ever larger catalogs of fundamental stellar parameters, coupled to the steady growth in the photometric database itself, almost entirely tied to traditional photoelectric photometry. With increasing frequency, the addition of new data has led to photometric calibration equations for individual stellar parameters that contain more terms of higher order, culminating in the unparalleled metallicity function of MA using 20 terms, driven primarily by the need to provide a universal function at all colors and abundances.

With the appearance of the catalog of VF, the potential exists to test if such complex polynomials have reached their limit, in the sense that the calibrations now supply less reliable abundance estimates than the photometric data allow, while introducing unnecessary scatter into larger statistical samples. To test this possibility and to see how the metallicity depends upon the primary indices of the $uvby$ photometric system, an expanded database for the analysis has been constructed. It is composed of a carefully transformed composite catalog of select high-dispersion spectroscopic abundances for 1801 stars on the metallicity scale of VF.  This catalog has then been cross-referenced with a merged catalog of high-precision $uvby$H$\beta$ photometry for over 35,000 stars. Simple empirical comparisons demonstrate that while [Fe/H] for F dwarfs is strongly dependent on $m_1$, with only a modest dependence on $c_1$, as expected from 30 years of stellar applications of the system, for hotter G dwarfs the pattern reverses and $c_1$ becomes the dominant index. As one moves to cooler G dwarfs and early K stars, the $c_1$ dominance continues. However, between $b-y$ = 0.50 and 0.58, a discontinuity appears such that stars with [Fe/H] $\geq$ +0.25 have $m_1$ and $c_1$ indices that classify them as subgiants, confirming an earlier result based upon a much smaller sample \citep{tw02}. The reversal in the sensitivity to $m_1$ and $c_1$ is traced, in part, to the metallicity sensitivity of the $b-y$ index; the reddening of $b-y$ at higher [Fe/H] shifts the star in the traditional two-color diagrams to a position of higher apparent luminosity/larger $c_1$ at a given $b-y$ and lower metallicity from $m_1$ relative to the standard relations. The combination of the reversal and the discontinuity also explains why the metallicity calibrations of MA and N04 produce growing scatter among the residuals in metallicity for the cooler stars.  Moreover, $b-y$ grows larger in a non-linear fashion for stars above solar metallicity, leading to an overestimate of the reddening for super-metal-rich stars from some standard intrinsic color relations. 

Given the increasing sensitivity of $b-y$ to rising [Fe/H], a potential solution that could weaken, if not remove, the effects of [Fe/H] on the reversal in sensitivity of $m_1$ and $c_1$ for cooler stars was tested using indices from synthetic spectra and the empirical data. Both approaches indicated that metallicity calibrations tied to H$\beta$ rather than $b-y$ have the potential to enhance the metallicity sensitivity of the $m_1$ index at all temperatures while weakening the luminosity effect through $c_1$. Keeping in mind that the sample of stars with a complete set of $uvby$H$\beta$ is about three-quarters the size of the $uvby$ sample, the prediction is borne out by defining a metallicity calibration at each H$\beta$ between 2.64 and 2.55, with at most five terms in the calibration. For stars with [Fe/H] $\geq$ $-$1.0, the typical scatter among the residuals ranges from 0.061 dex at the hot end to a maximum of 0.085 dex among the cooler stars. Moreover, the use of the specialized calibrations with fewer terms should lead to smaller errors tied to uncertainties in both reddening determination and photometric scatter.

Do the new calibrations resolve the problem of the discontinuity/confusion among super-metal-rich dwarfs and the subgiants? Unfortunately, only a small fraction of the stars of interest have H$\beta$ data, so a definitive conclusion isn't possible at present. Of the eight super-metal-rich dwarfs in the discontinuity, four have all the required indices. The mean difference in [Fe/H] for these stars, in the sense (SPEC - PH), with the correction derived above included is +0.011 $\pm$ 0.094.  The sample for the subgiants is even smaller; three stars have mean residuals of  $-$0.005 $\pm$ 0.130. For better insight into this discontinuity and to expand the  successful application of the H$\beta$ system to cooler dwarfs and subgiants, H$\beta$ observations of additional cooler stars in the solar neighborhood could prove informative.

Finally, assuming the calibrations defined above prove valid for all metallicities, including super-metal-rich dwarfs, what potential applications exist beyond simply improving the photometric abundances of nearby stars? The obvious transition that has dominated all photometry over the last 20 years has been the switch to CCD detectors of ever-larger format and improved sensitivity. By contrast, all the photometric data discussed in this investigation and those noted in the introduction are the product of traditional photoelectric technology. With few exceptions \citep{at00}, CCD investigations with intermediate and narrow-band photometry have focused on star clusters where the areal coverage afforded by CCD's can be maximized, especially if photoelectric observations within the cluster can be accessed for calibration purposes. While the value of these studies has been the demonstration that CCD $uvbyCa$H$\beta$ photometry competitive with traditional photoelectric approaches is possible if care is taken to obtain adequate standard star data, the use of CCD's for all-sky photometry on these systems has virtually disappeared.

As exemplified by the followup work of VF and N04, there can be enormous potential payoff in detailed analysis of well-defined subsamples of stars selected from large surveys tied to either low-resolution spectroscopy or broad-band photometry. Telescopes in the 1-meter to 2-meter class may remain adequate for efficient higher resolution studies of samples containing on the order of $10^4$ stars in the magnitude range between $V$ = 6 and 10.  However, the availability of all-sky samples of broad-band photometry extending to significantly fainter magnitudes opens the possibility of probing the stellar dwarf population well beyond the 50 pc radius around the sun, as well as identifying statistically viable samples of extreme stellar populations that are rarely found in current studies, e.g., stars with [Fe/H] $\ge$ +0.5 or subgiant CH stars \citep{bo74}. Because of the observational overhead in doing detailed analysis of stars in the $V$ = 10 to 15 magnitude range, i.e., either fewer stars and/or larger telescopes, an intermediate step in the selection process between broad-band photometry and spectroscopy based upon more specialized photometric systems could be extremely cost effective. With a 1-m telescope, photometry good to the canonical 0.01 mag accuracy is achievable for the full array of $uvbyCa$H$\beta$ indices down to $V$ = 17 with only modest effort. As demonstrated by this investigation, the system has many capabilities for cooler stars, including the ability to separate cool dwarfs from giants, to generate individual reddening and metallicity estimates, and, potentially, to allow identification of super-metal-rich dwarfs. Given the statistical distortion created by the volume effect on a magnitude-limited sample, isolating cool dwarfs from distant giants will be critical in minimizing the followup time wasted observing unwanted interlopers in the sample; CCD survey photometry on the $uvbyCa$H$\beta$ system offers a feasible intermediate solution.

\acknowledgements
Extensive use was made of the SIMBAD database, operating at CDS, Strasbourg, France. LCV gratefully acknowledges the support of a Goldwater Scholarship during the period of this research.  The groundwork for this project was begun as part of the CSUURE REU program at Mount Laguna Observatory; the authors and Lindsay Mayer express their appreciation to the REU director, Eric Sandquist and MLO's director, Paul Etzel. The clarity and accuracy of the paper have been improved thanks to a careful reading by the referee, Howard Bond.

\clearpage
\figcaption[f01.eps]{Histograms of the metallicity distributions for, from top to bottom, all stars in the composite spectroscopic catalog, stars with $uvby$ data, and stars with $uvby$H$\beta$ photometry. \label{f01}} 

\figcaption[f02.eps]{Histograms of the $b-y$ distributions for (upper histogram) all stars with $b-y$ photometry and (lower histogram) stars with both $b-y$ and H$\beta$ photometry. \label{f02}}

\figcaption[f03.eps]{Comparisons between the spectroscopic abundances and the photometric abundance estimates using the calibration from MA (circles) and from N04 (triangles) as a function of color. Filled symbols illustrate the average residuals while open symbols show the dispersion among the residuals. The stars show the dispersion among the spectroscopic abundances with color. \label{f03}}

\figcaption[f04.eps]{Residuals in [Fe/H], in the sense (SPEC-PHOT), from the calibration of MA for stars with $b-y$ between 0.39 and 0.50. The line is the mean relation through the points. \label{f04}}

\figcaption[f05.eps]{Spectroscopic abundances for stars with $b-y$ between 0.300 and 0.329 as a function of $m_1$. \label{f05}}

\figcaption[f06.eps]{Spectroscopic abundances for stars with $b-y$ between 0.300 and 0.329 as a function of $c_1$. \label{f06}}

\figcaption[f07.eps]{ Same as Fig. 5 for stars between $b-y$ = 0.44 and 0.469. \label{f07}}

\figcaption[f08.eps]{Same as Fig. 6 for stars between $b-y$ = 0.44 and 0.469. \label{f08}}

\figcaption[f09.eps]{Same as Fig. 5 for stars between $b-y$ = 0.510 and 0.549. Filled circles are unevolved main sequence stars; open circles are subgiants and giants. \label{f09}}

\figcaption[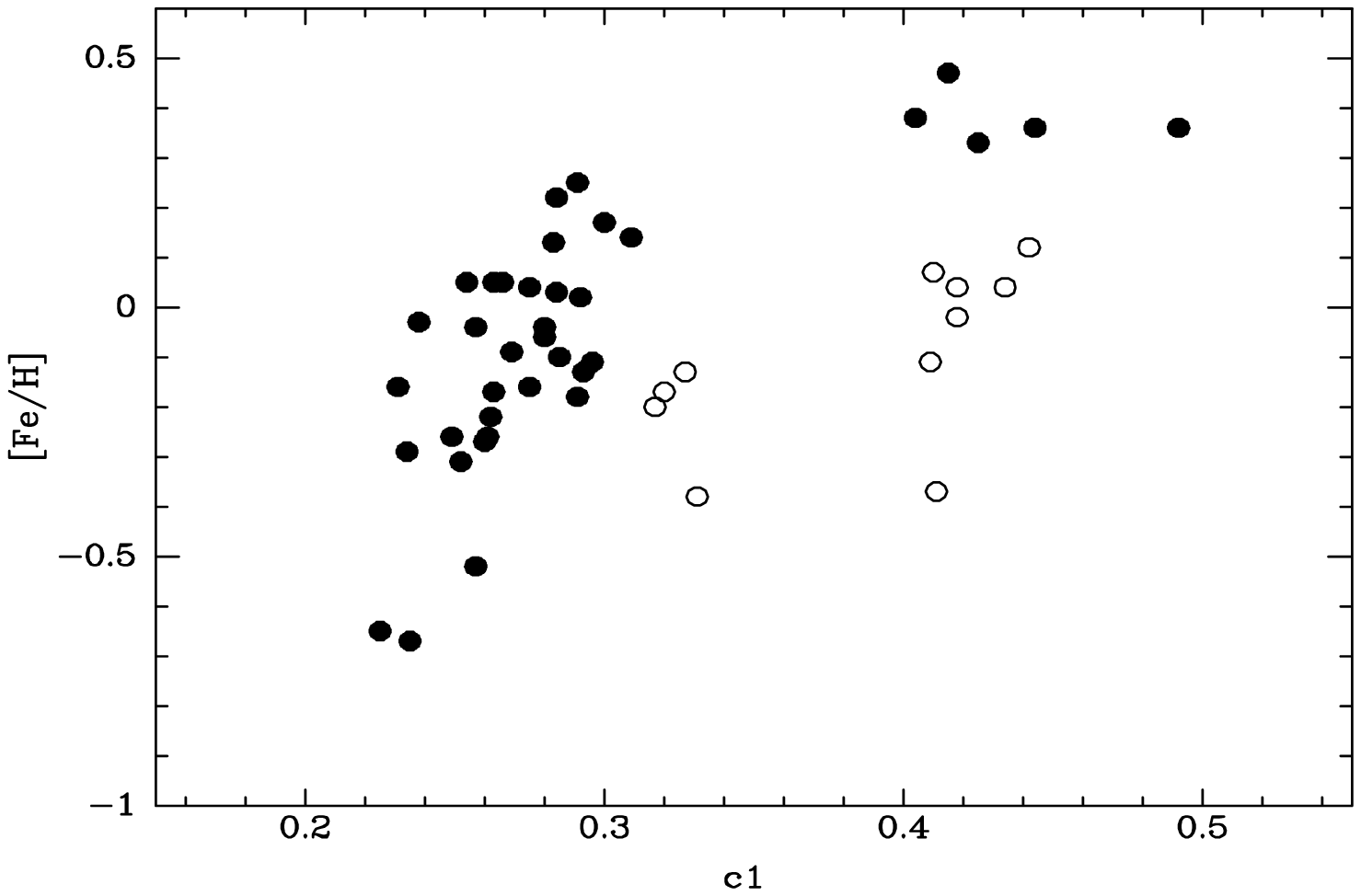]{Same as Fig. 6 for stars between $b-y$ = 0.510 and 0.549. Filled circles are unevolved main sequence stars; open circles are subgiants and giants. \label{f10}}

\figcaption[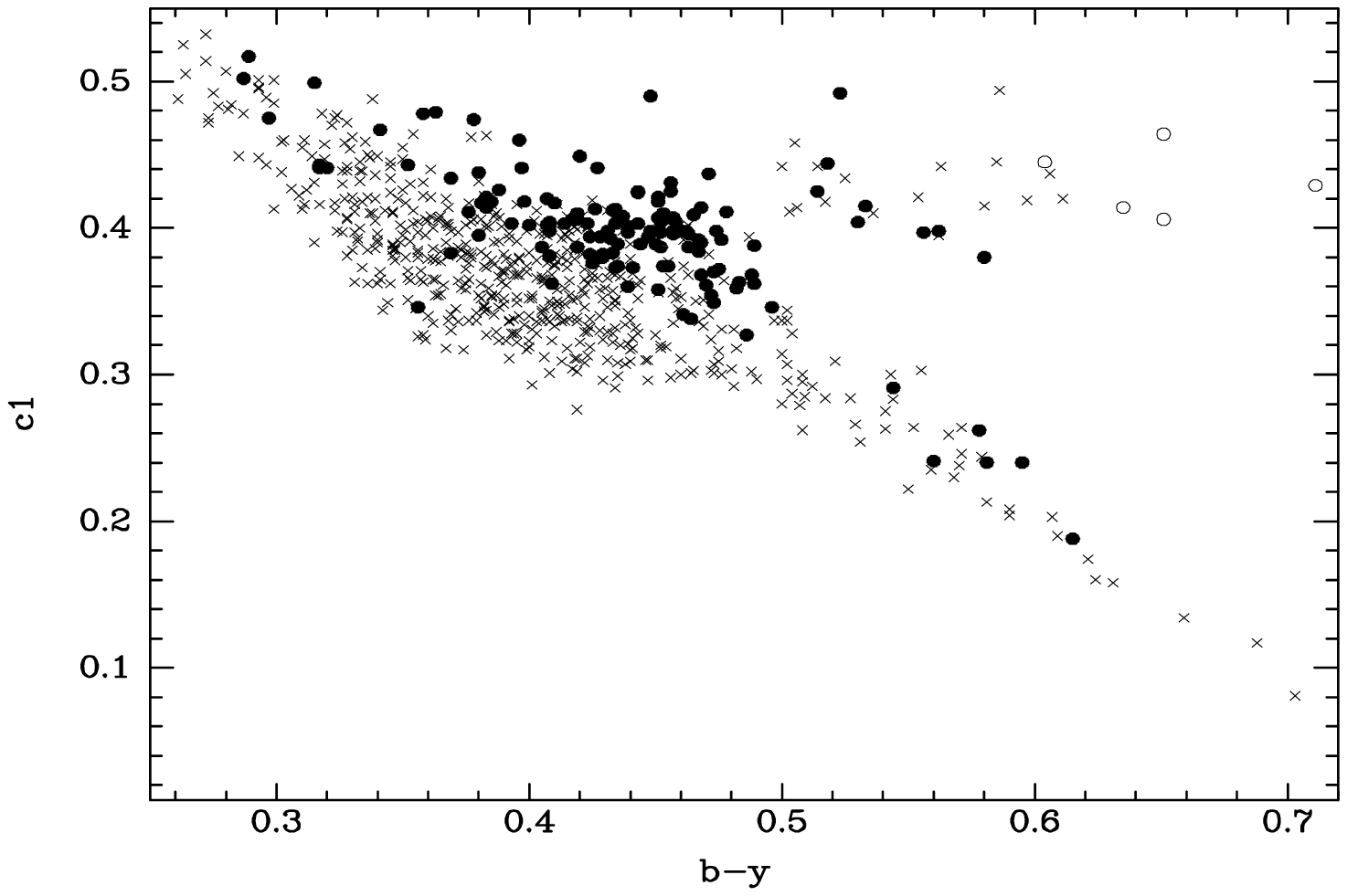]{The $c_1,b-y$ diagram for stars with spectroscopic abundances between 0.0 and +0.24 (crosses) and [Fe/H] $\geq$ +0.25 (filled circles). Evolved (subgiant/giant) stars with [Fe/H] $\geq$ +0.25 are plotted as open circles.  \label{f11}}

\figcaption[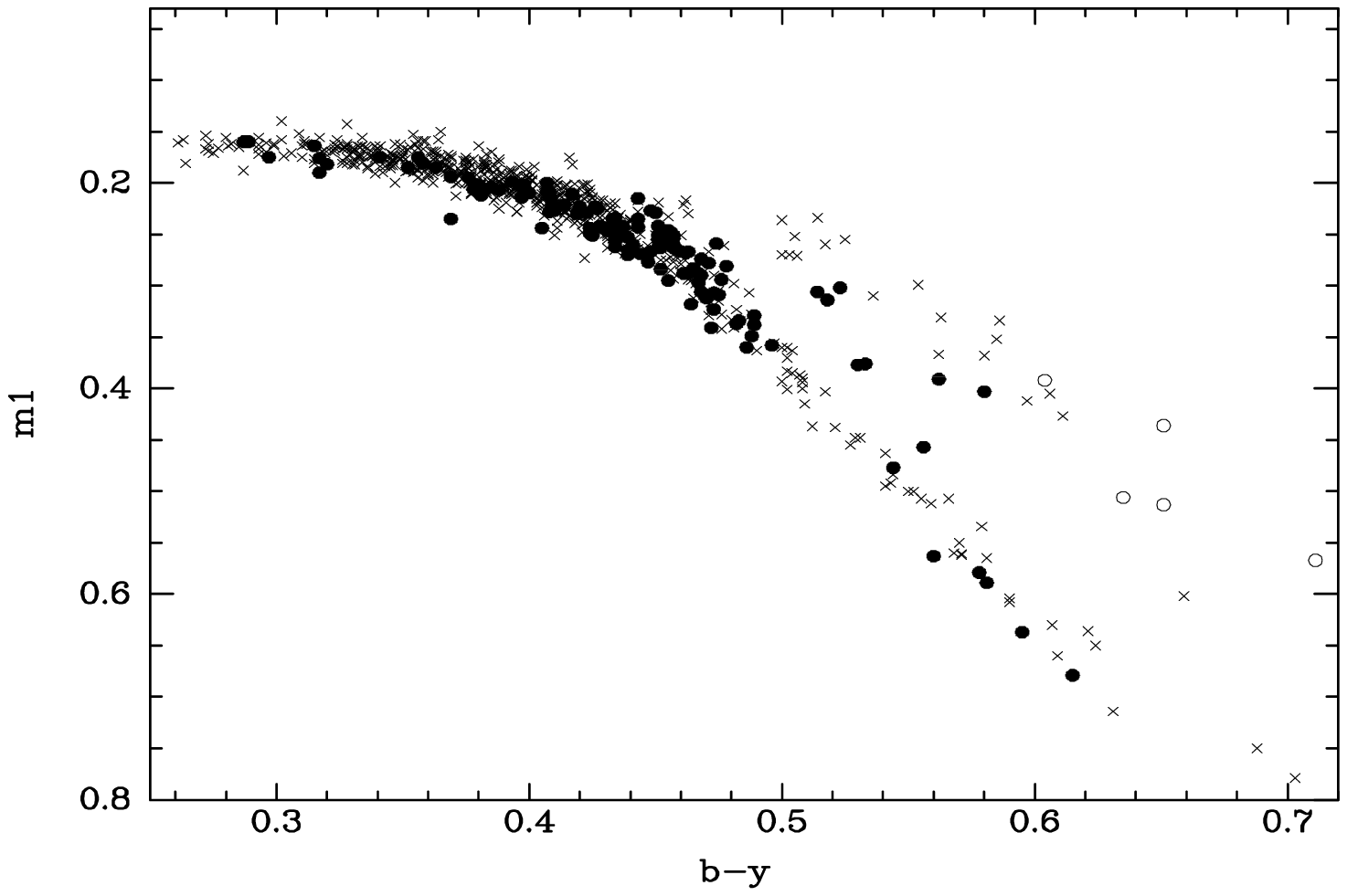]{Same as Fig. 11 for $m_1, b-y$. \label{f12}}

\figcaption[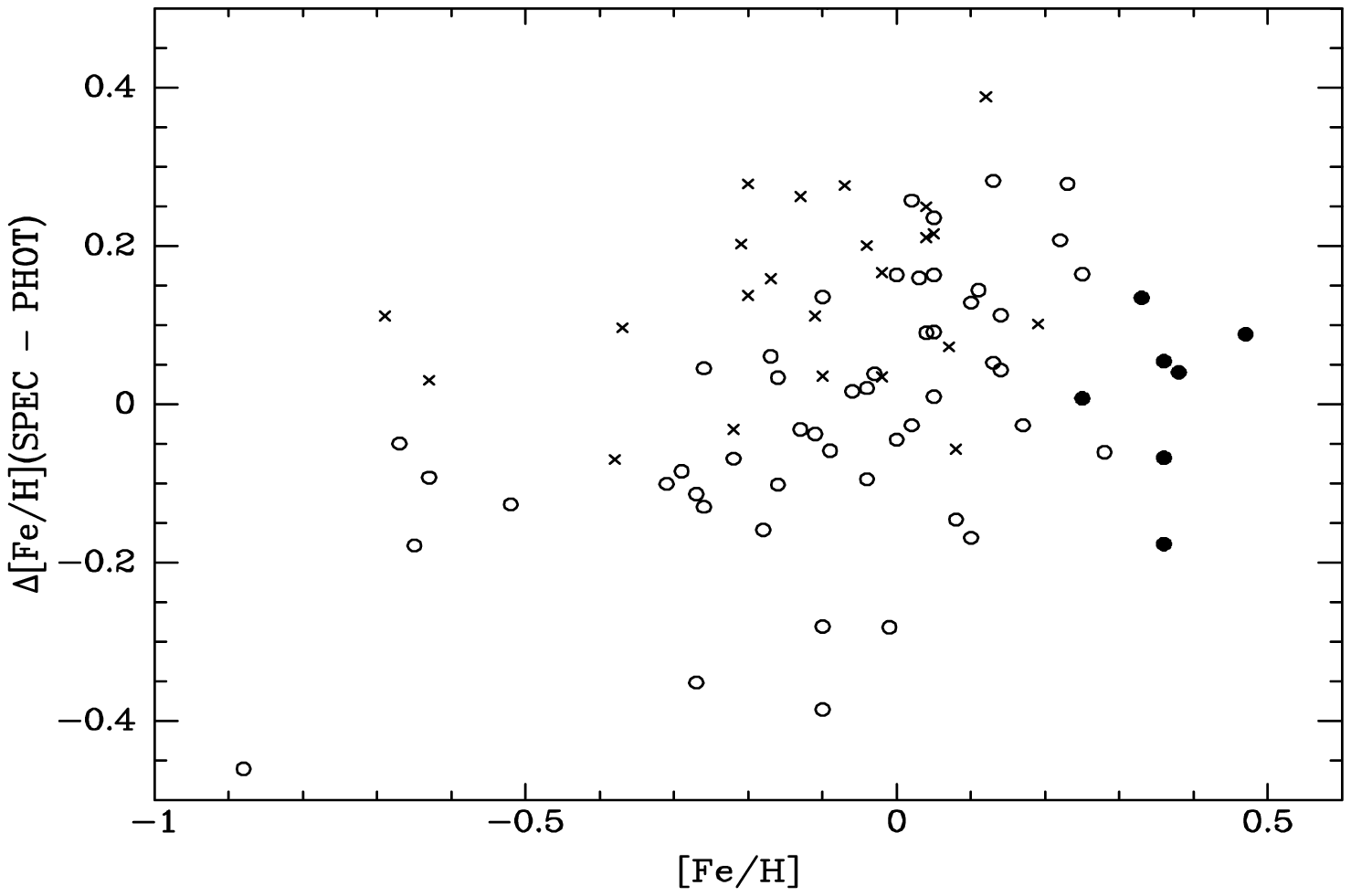]{Residuals in [Fe/H], in the sense (SPEC-PHOT), for cool dwarfs (circles) and evolved stars (crosses) using the calibration of MA. The filled circles are the super-metal rich dwarfs that overlap in $c_1$ with the evolved stars. \label{f13}}

\figcaption[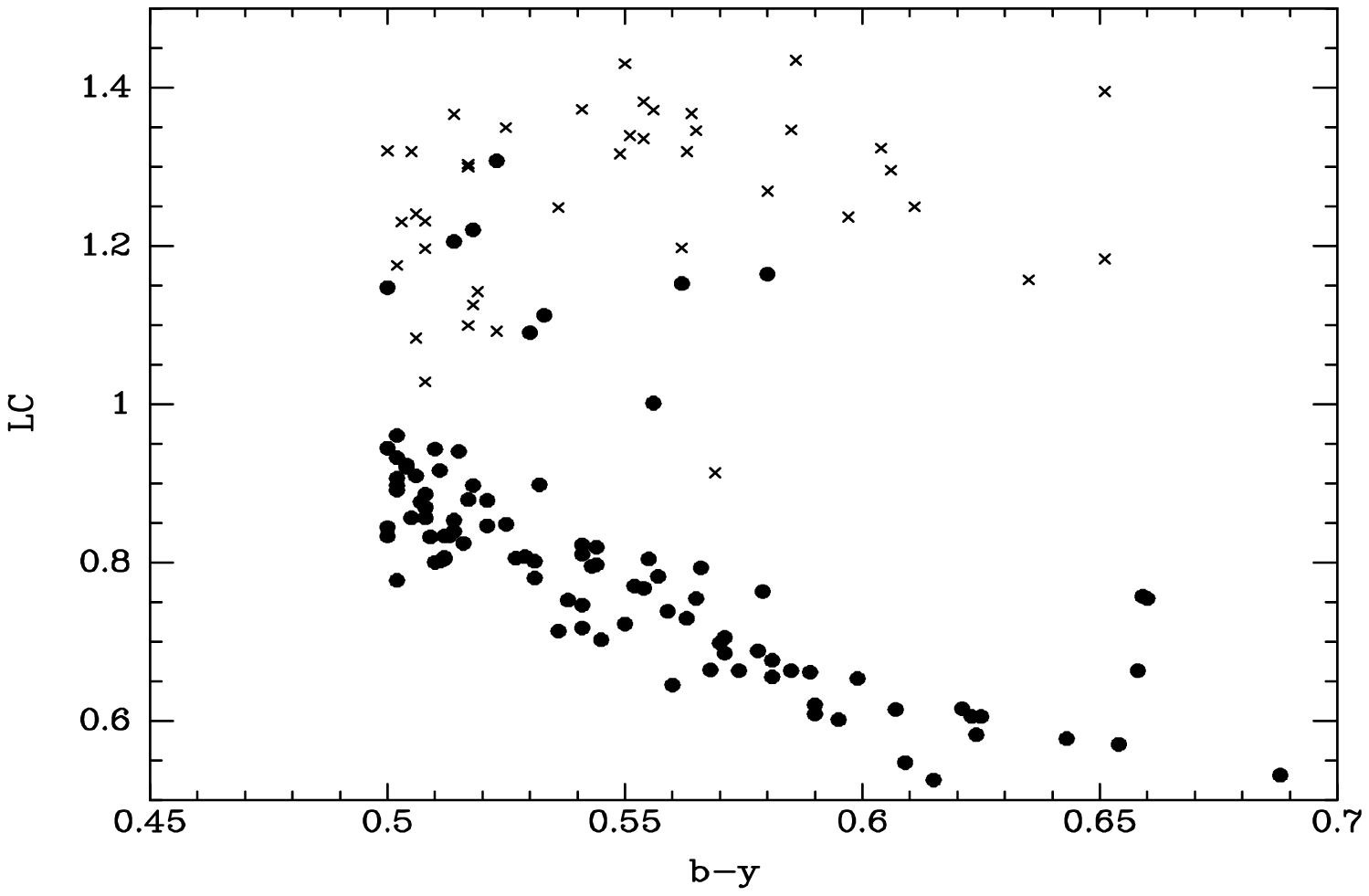]{The luminosity class parameter, LC, as a function of $b-y$ for stars with log g $\geq$ 4.2 (filled circles) and log g $<$ 4.2 (crosses). \label{f14}}

\figcaption[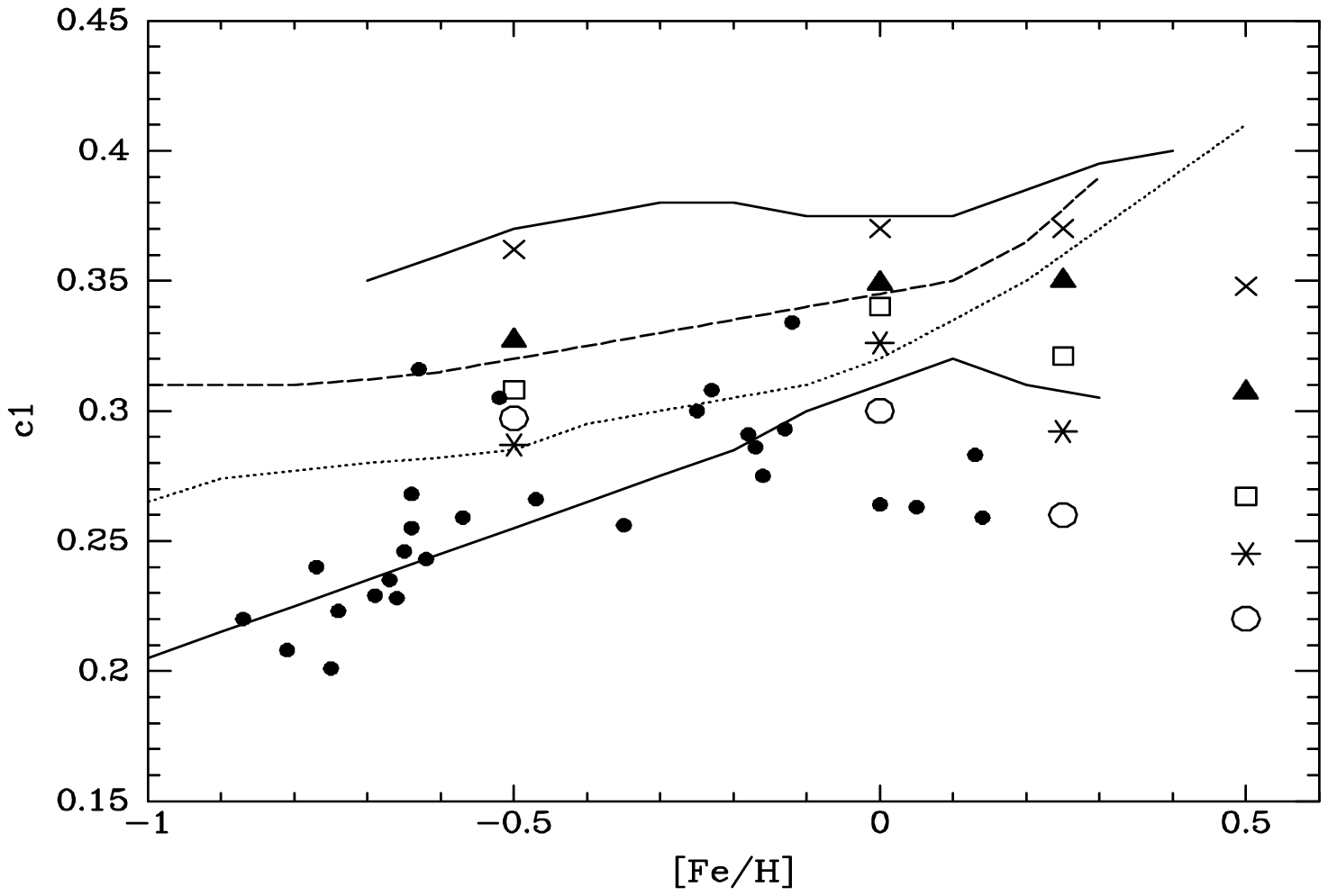]{The correlation of $c_1$ with [Fe/H] for data sorted by H$\beta$. Trends based upon synthetic spectra for H$\beta$ = 2.56, 2.58, 2.60, 2.62, and 2.64 defined by open circles, stars, open squares, filled triangles, and crosses, respectively. Curves from bottom to top are the mean relations through the data for H$\beta$ = 2.56, 2.58, 2.60, and 2.62, respectively. Filled circles are all stars observed with H$\beta$ between 2.35 and 2.49. \label{f15}}

\figcaption[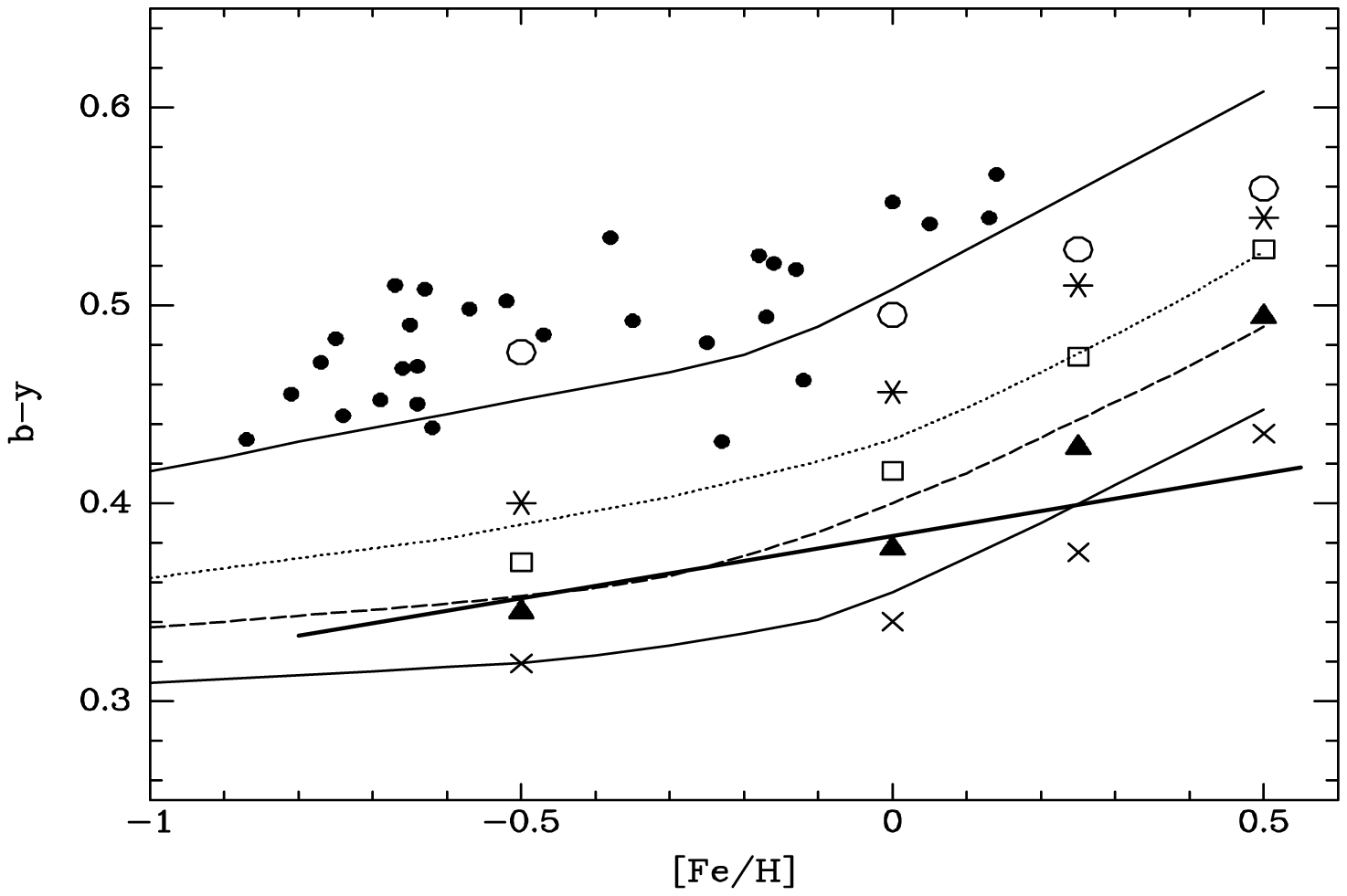]{The correlation of $m_1$ with [Fe/H] for data sorted by H$\beta$. Symbols have the same meaning as in Fig. 15. Curves from top to bottom are the mean relations through the data for H$\beta$ = 2.56, 2.58, 2.60, and 2.62, respectively. Filled circles are all stars observed with H$\beta$ between 2.35 and 2.49. The solid, thick straight line is the predicted trend for H$\beta$ = 2.60 from the average of the relations of \citet{o88} and \citet{n88}. \label{f16}}

\figcaption[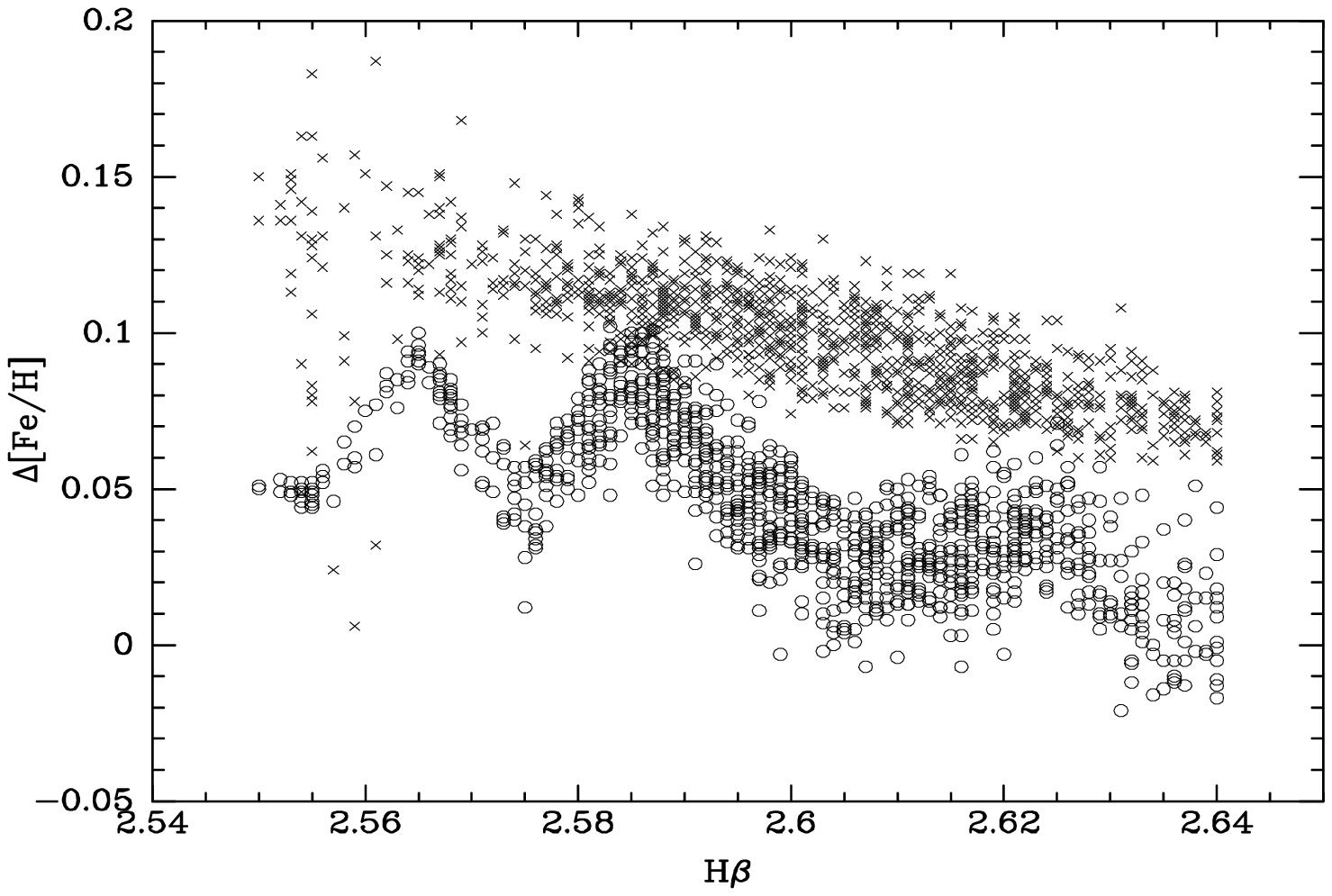]{Plot of the residuals between the true photometric abundance and the photometric abundance if a reddening of E$(B-V)$ = 0.020 is applied to the stars without correction. Crosses show the results for the calibration of MA, while the open circles use the relations of Table 2. \label{f17}}

%\end{document}
\newpage
\plotone{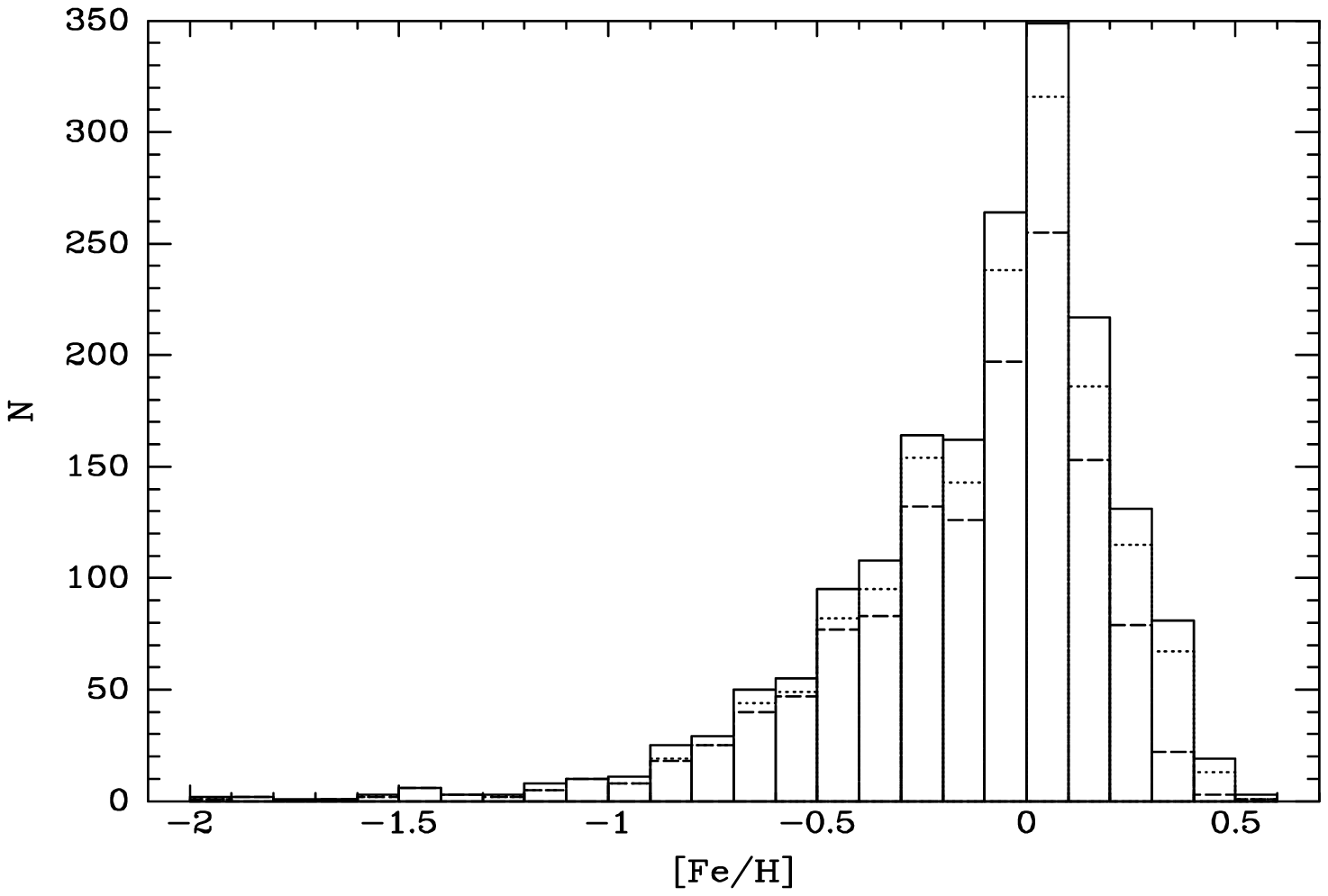}
\plotone{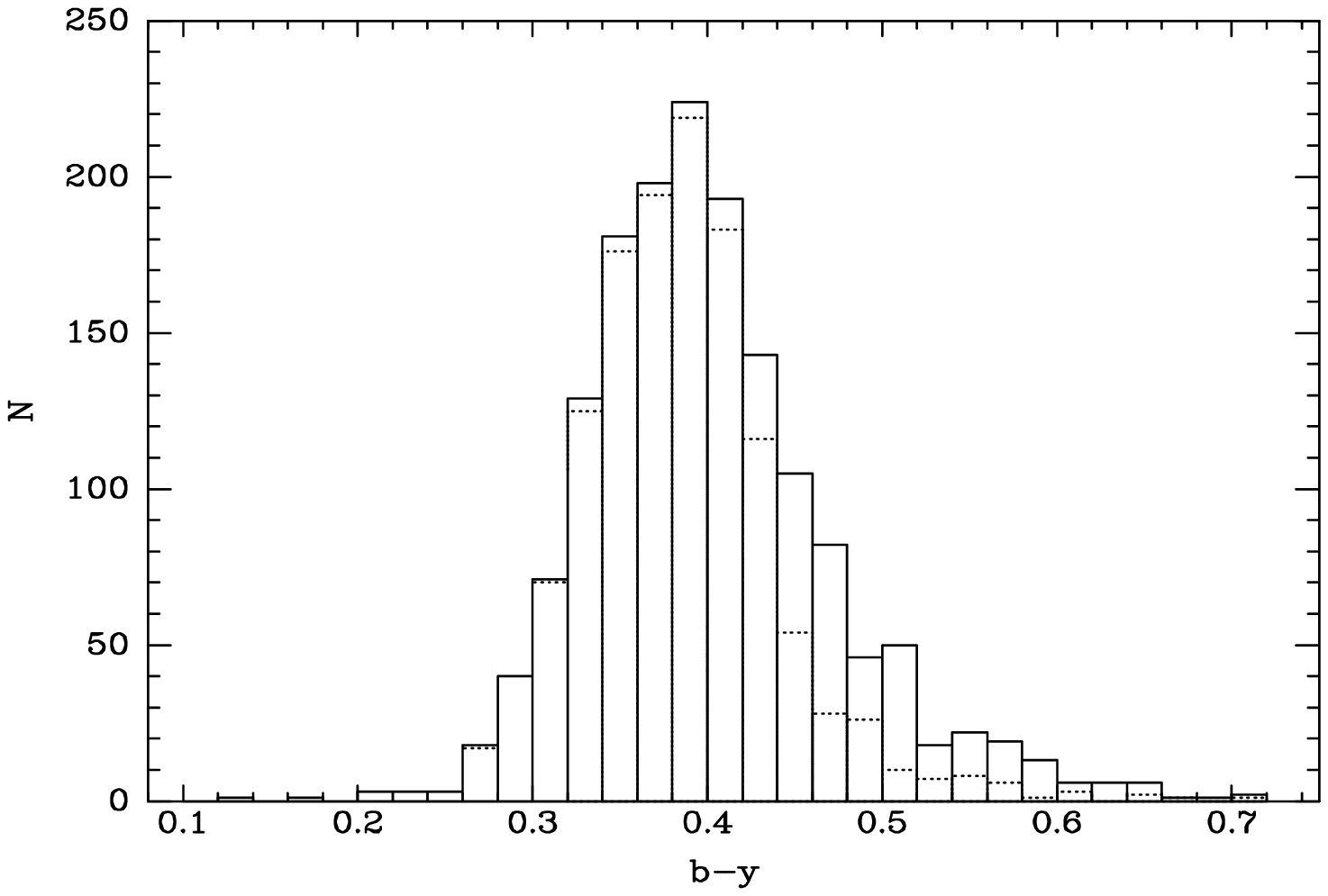}
\plotone{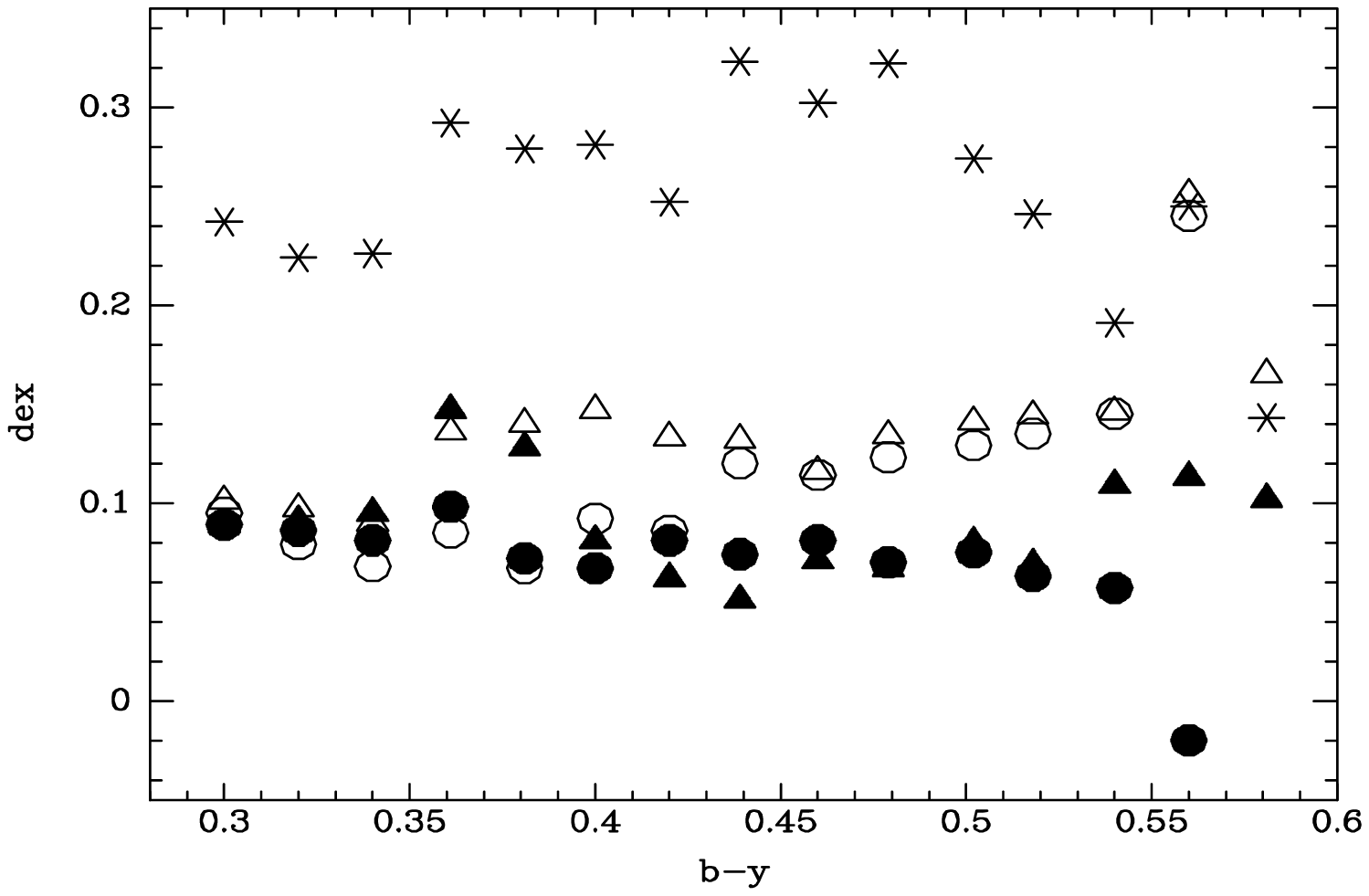}
\plotone{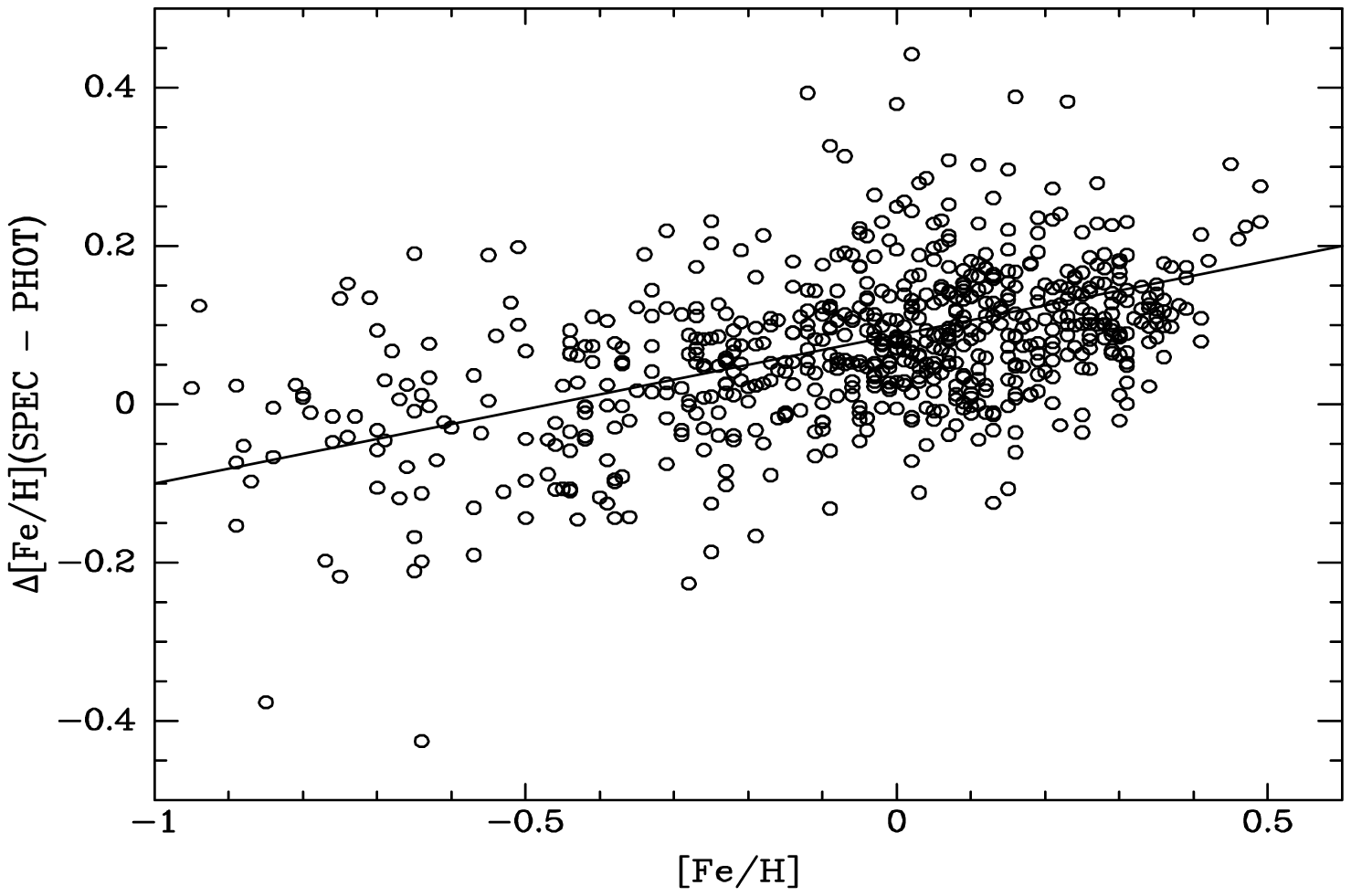}
\newpage
\plotone{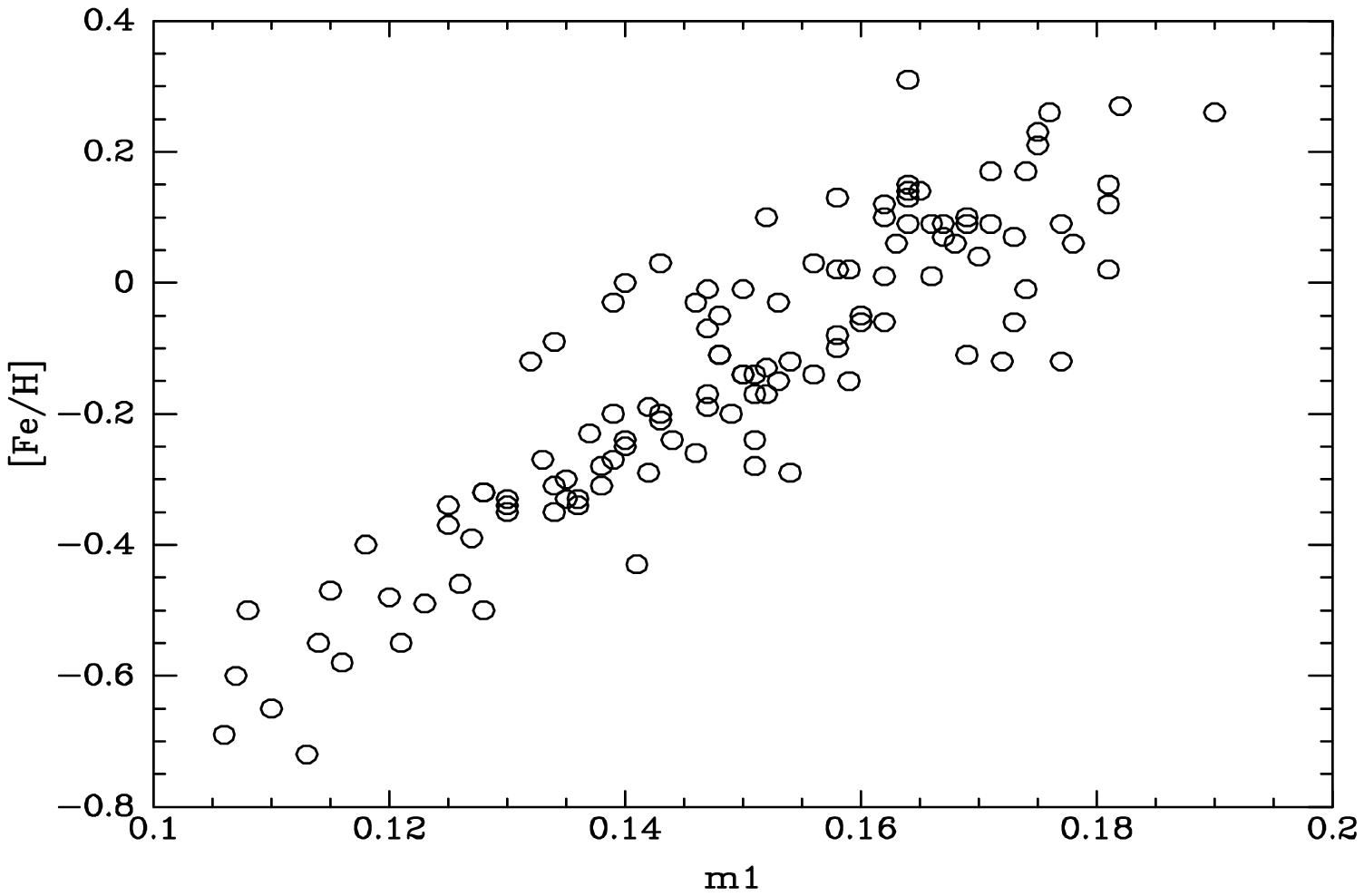}
\plotone{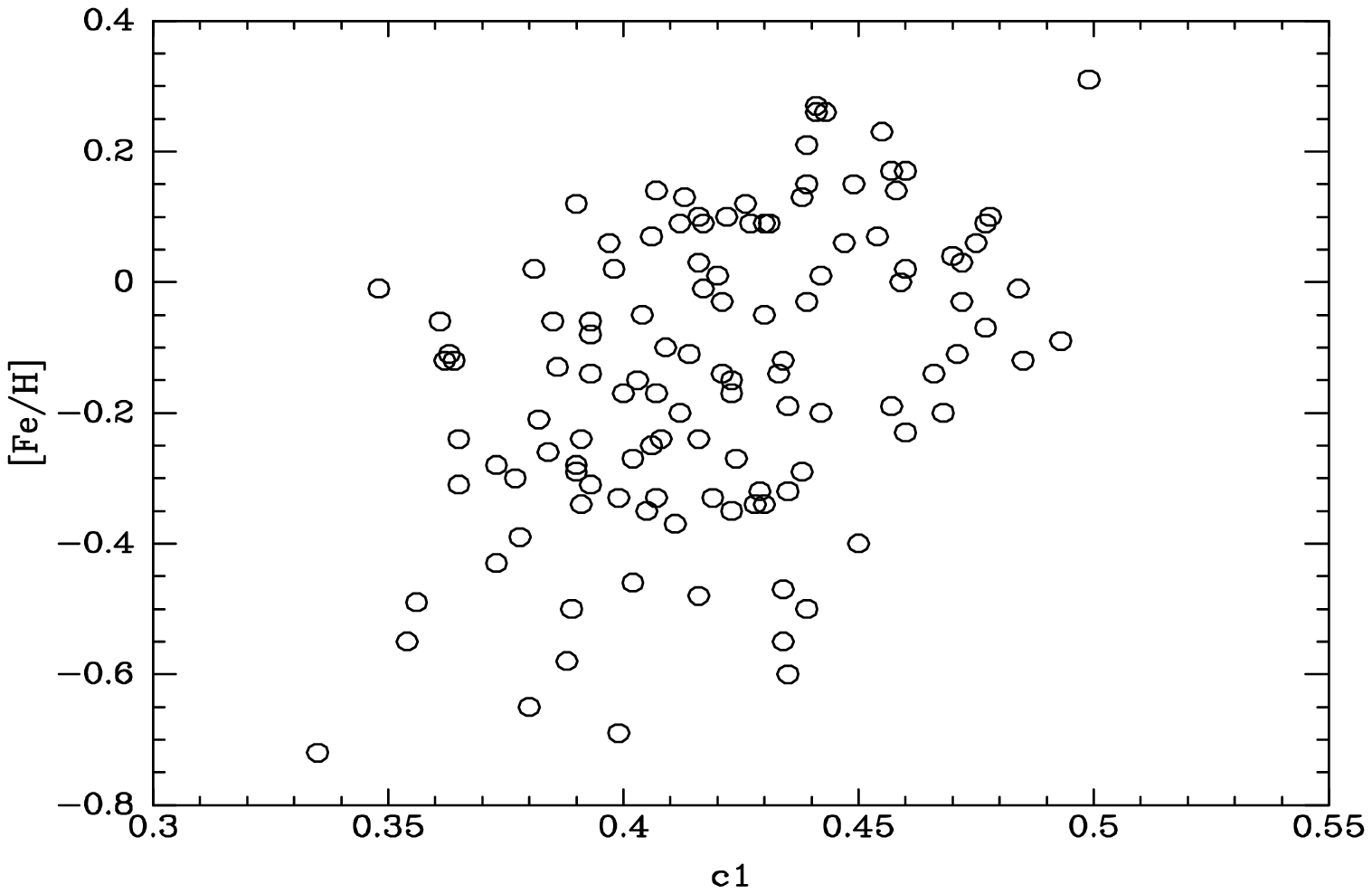}
\plotone{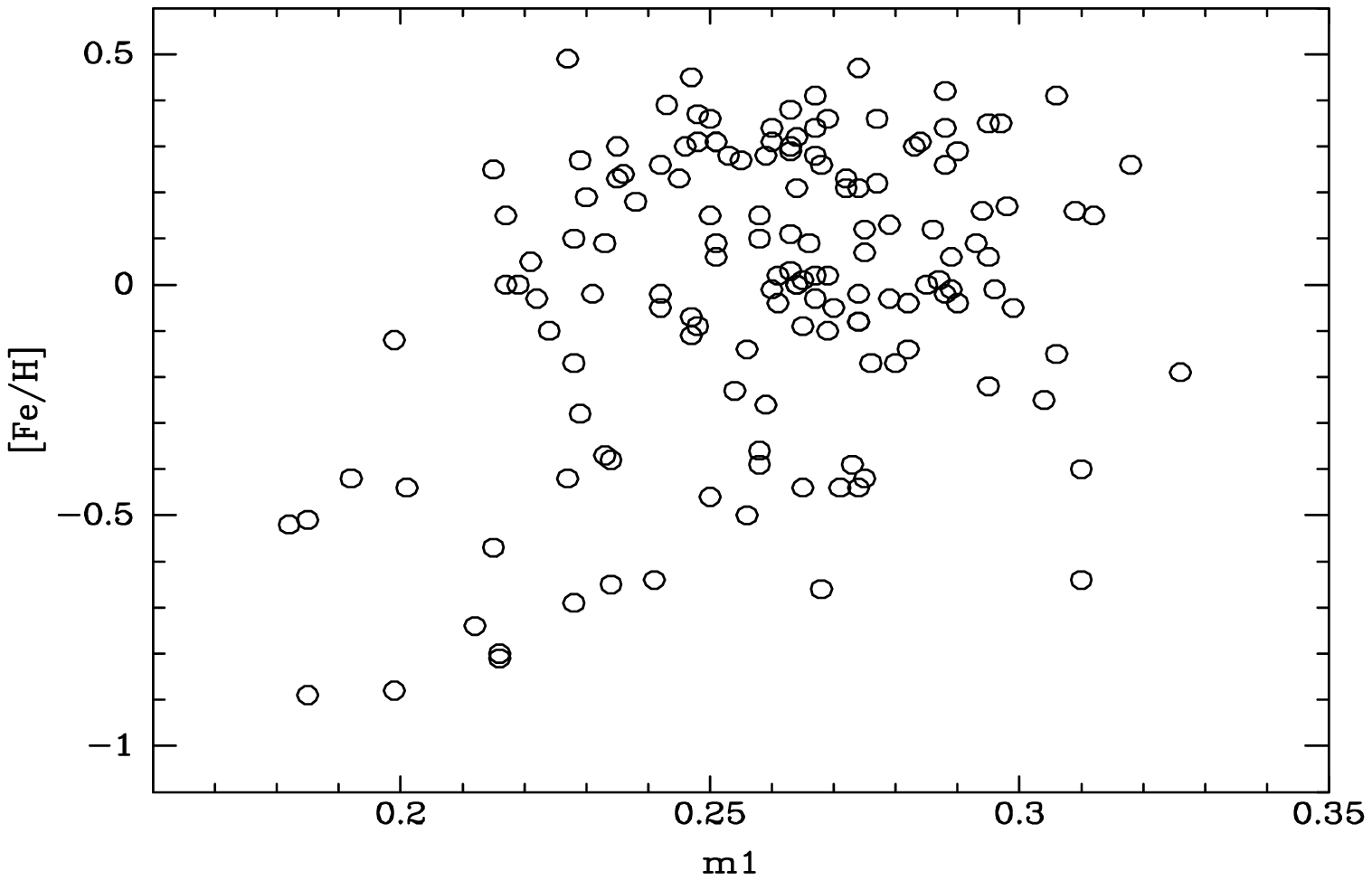}
\plotone{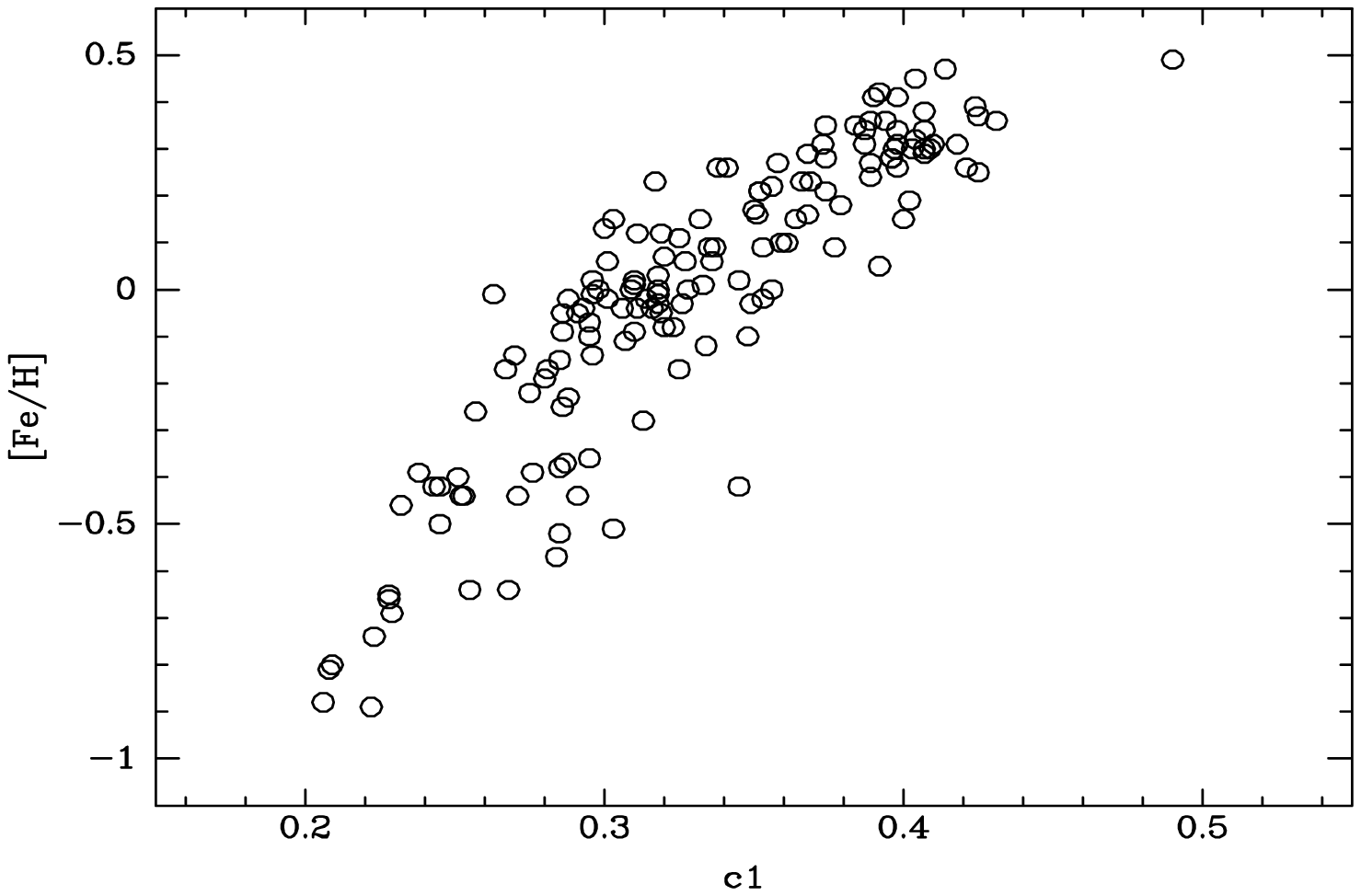}
\newpage
\plotone{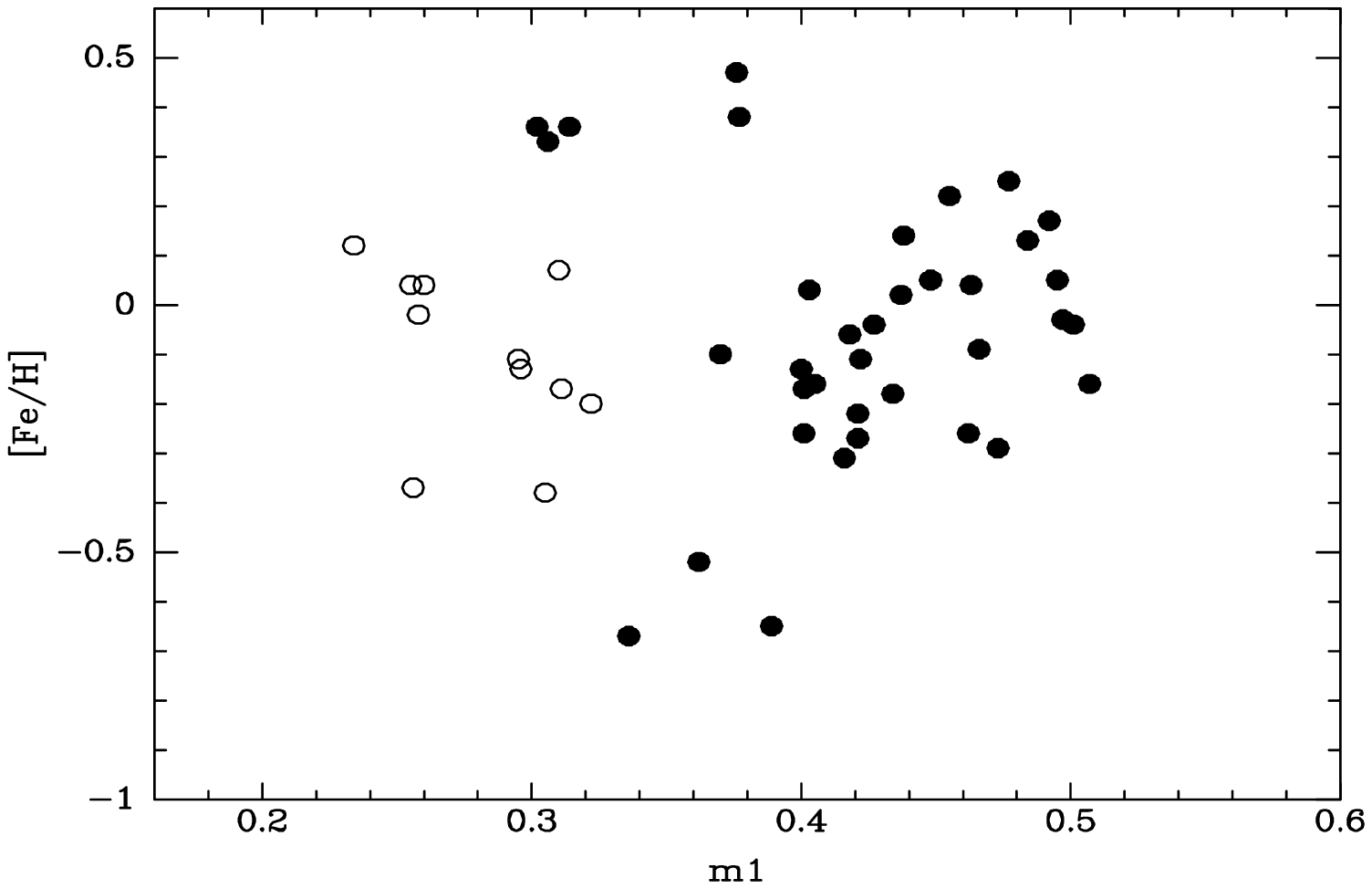}
\plotone{f10.eps}
\plotone{f11.eps}
\plotone{f12.eps}
\newpage
\plotone{f13.eps}
\plotone{f14.eps}
\plotone{f15.eps}
\plotone{f16.eps}
\newpage
\plotone{f17.eps}
\newpage
%\documentclass[preprint]{aastex}
%\begin{document}
%\pagestyle{empty}
%\documentstyle{article}
%\begin{document}
%\pagestyle{empty}
%\s\iptsize
\begin{deluxetable}{lrrrrrrrrrrrrrrc}
%\rotate
%\tabletypesize\small
\tablenum{1}
\tablecolumns{8}
\tablewidth{0pc}
\tablecaption{Spectroscopic Catalog Summary Information}
\tablehead{
\colhead{Spectroscopic Source}     & 
\colhead{$N_{tr}$}   & 
\colhead{$N_{cat}$}  & 
\colhead{$\sigma$}    & 
\colhead{$a$}   & 
\colhead{$b$}   & 
\colhead{$c$}  &
\colhead{$d$} } 
\startdata
Boesgaard and Friel (1990) &    18 &  70& 0.049   &  0.718  &   0.000 &    0.000 & 0.038 \cr
Edvaardsson et al (1993)   &   145 & 188& 0.046   &  1.000  &  -0.510 &  -0.253 & 3.044 \cr
Furhmann (1998)            &    48 & 50 & 0.029   &  1.000  &  0.480  &  -0.091 & -1.427 \cr
Tomkin and Lambert (1999)  &   16 &  31 & 0.052   &  1.000  & -1.812  & -0.076 & 7.039  \cr
Chen et al. (2000)         &   60 &   90& 0.052   &  1.000  & -1.637  & -0.149 &  6.881\cr
Gonzalez et al. (2001)     &   29 &  29 & 0.025   &  1.000  & -1.320  & -0.099 &  5.370\cr
Gaidos and Gonzalez (2002) &   22 &   33& 0.025   &  1.000  &  -0.744 &  -0.103 &  3.289\cr
Qui et al. (2002)          &  19  &   24& 0.048   &  1.000  &  -0.518 &  -0.126 &  2.380\cr
Sadakane et al. (2002      &  12  &   12& 0.025   &  0.940  &   0.000 &   -0.084 &  0.409 \cr
Laws et al. (2003)         &  31  &   31& 0.026   &  1.000  &  -0.556 &   -0.069 &  2.367\cr
Reddy et al. (2003)        &  34  &  181& 0.026   &  1.000  & -2.728  &  -0.069 & 10.640\cr
Allende Prieto et al. (2004)& 95  &  103& 0.050   &  1.000  &   0.000 &    0.000 &  0.046\cr
Fuhrmann (2004)            & 111  &  132& 0.042   &  1.000  &   0.000 &   -0.073 &  0.314\cr
Galeev et al. (2004)       &  14  &   15& 0.052   &  0.900  &   0.000 &    0.081 & -0.362\cr
Mishenina et al. (2004)    & 124  &  173& 0.053   &  1.000  &  -1.758 &   -0.077&  6.977 \cr
Bensby et al. (2005)       &  77  &  102& 0.026   &  0.941  &    0.000 &  -0.038 &  0.160\cr
Bonfils et al. (2005)      &  17  &   21& 0.052   &  0.892  &   -0.736 &  -0.056 &  3.052 \cr
Huang et al. (2005)        &  22  &   22& 0.039   &  0.903  &   -0.741 &   -0.173 &  3.507\cr
Takeda et al. (2005)       & 155  &  160& 0.037   &  1.000  &   0.000  &   -0.148 &  0.618\cr
King and Schuler (2005)    &   3  &    7& 0.040   &  1.000  &    0.000  &   0.000  &  0.043\cr
Valenti and Fischer (2005) &1039  & 1039& 0.025   & \nodata &  \nodata  &  \nodata & \nodata   \cr
Chen and Zao (2006)        & 34   &   34& 0.043   &  1.000  &  -5.800   & -0.155 & 22.485 \cr
Gilli et al. (2006)        & 161  &  193& 0.030   &  1.000  &  -1.172  &  -0.075  &  4.714 \cr
Luck and Heiter (2006)     & 169  &  217& 0.043   &  0.876  &  0.279  & -0.069  & -0.736 \cr
Reddy et al. (2006)        & 39   &  175& 0.061   &  1.000  &  0.000  & -0.138  &  0.596 \cr
Bond et al. (2006)         & 135  &  136& 0.044   &  1.000  & -1.359  & -0.237 &  6.195 \cr
\enddata
\end{deluxetable}
%\end{document}

\newpage
%\documentclass[preprint]{aastex}
%\begin{document}
%\pagestyle{empty}
%\documentstyle{article}
%\begin{document}
%\pagestyle{empty}
\begin{deluxetable}{lrrrrrrrrrrrrrrc}
%\rotate
%\tabletypesize\small
\tablenum{2}
\tablecolumns{8}
\tablewidth{0pc}
\tablecaption{Polynomial Coefficients for the [Fe/H], H$\beta$ Calibration}
\tablehead{
\colhead{H$\beta$}     & 
\colhead{N}     & 
\colhead{$\sigma$}     & 
\colhead{$a$}     & 
\colhead{$b$}     & 
\colhead{$c$}     & 
\colhead{$d$}  &
\colhead{$e$} } 
\startdata 
  2.635 &  75& 0.068 & -4.375 & 30.934& -73.762 &  1.627 &  1.869 \cr
  2.625 & 127& 0.064 & -2.980 & 22.825& -45.140 &  1.058 & \nodata \cr
  2.615 & 190& 0.061 & -3.471 & 24.465& -50.476 &  1.925 & \nodata \cr
  2.605 & 181& 0.062 & -2.835 & 17.484& -31.388 &  1.820 & \nodata \cr
  2.595 & 198& 0.065 & -2.644 & 18.586& -32.043 &  2.636 & -1.750 \cr
  2.585 & 165& 0.068 & -1.928 & 16.668& -24.974 &  3.376 & -3.964 \cr
  2.575 & 75 & 0.075 & -2.383 & 12.638& -17.934 &  3.455 & -1.761 \cr
  2.565 & 49 & 0.085 & -0.691 &  8.876&  -7.806 &  3.654 & -5.170 \cr
  2.555 & 36 & 0.079 & -1.275 &  5.664&  -2.911 &  2.538 & -2.461 \cr
\enddata
\end{deluxetable}
%\end{document}

\end{document}